\documentclass[]{pasj02} 
\usepackage[switch,mathlines]{lineno} 
\usepackage{bm}
\usepackage{comment}
\usepackage{amsmath}
\usepackage{color}
\usepackage[whole]{bxcjkjatype}

\jyear{2024}
\Received{}
\Accepted{}

\begin{document} 

\title{Discovery of an Eccentric Hot Super-Jupiter Leaving the Transiting Geometry of the Early-A-type star TOI-1355}

\author{Noriharu \textsc{Watanabe}\altaffilmark{1}\orcid{0000-0002-7522-8195}}
\email{n-watanabe@g.ecc.u-tokyo.ac.jp}
\author{Norio \textsc{Narita}\altaffilmark{2,3,4}\orcid{0000-0001-8511-2981}}
\author{Akihiko \textsc{Fukui}\altaffilmark{2,4}\orcid{0000-0002-4909-5763}}
\author{Bun'ei \textsc{Sato}\altaffilmark{5}\orcid{0000-0001-8033-5633}}
\author{Keisuke \textsc{Isogai}\altaffilmark{1,6}\orcid{0000-0002-6480-3799}}
\author{John \textsc{H. Livingston}\altaffilmark{3,7,8}\orcid{0000-0002-4881-3620}}
\author{Jerome \textsc{P. de Leon}\altaffilmark{2}\orcid{0000-0002-6424-3410}}
\author{Daniel \textsc{Huber}\altaffilmark{9}\orcid{0000-0001-8832-4488}}
\author{Yugo \textsc{Kawai}\altaffilmark{1}\orcid{0000-0002-0488-6297}}
\author{Yuya \textsc{Hayashi}\altaffilmark{1}\orcid{0000-0001-8877-0242}}
\author{Masashi \textsc{Omiya}\altaffilmark{3,7}}
\author{Hideyuki \textsc{Izumiura}\altaffilmark{10}\orcid{0000-0002-8435-2569}}
\author{Akito \textsc{Tajitsu}\altaffilmark{10}\orcid{0000-0001-8813-9338}}
\author{Keivan G.\ \textsc{Stassun}\altaffilmark{11}\orcid{0000-0002-3481-9052}}
\author{Eric \textsc{Girardin}\altaffilmark{12}\orcid{0000-0002-5443-3640}}
\author{Giuseppe \textsc{Marino}\altaffilmark{13,14}\orcid{0000-0001-8134-0389}} 
\author{Antonio \textsc{Frasca}\altaffilmark{14}\orcid{0000-0002-0474-0896}}
\author{Giovanni \textsc{Catanzaro}\altaffilmark{14}\orcid{0000-0003-4337-8612}}
\author{Javier \textsc{Alonso-Santiago}\altaffilmark{14}\orcid{0000-0001-9707-3107}}
\author{Manfred \textsc{Raetz}\altaffilmark{15}\orcid{0000-0002-2190-3319}}
\author{Stephanie \textsc{Striegel}\altaffilmark{16,17}\orcid{0009-0008-5145-0446}}
\author{Nobuhiko \textsc{Kusakabe}\altaffilmark{3,7,18}\orcid{0000-0001-9194-1268}}
\author{Motohide \textsc{Tamura}\altaffilmark{3,19}\orcid{0000-0002-6510-0681}}

\altaffiltext{1}{Department of Multi-Disciplinary Sciences, Graduate School of Arts and Sciences, The University of Tokyo, 3-8-1 Komaba, Meguro, Tokyo 153-8902, Japan}
\altaffiltext{2}{Komaba Institute for Science, The University of Tokyo, 3-8-1 Komaba, Meguro, Tokyo 153-8902, Japan}
\altaffiltext{3}{Astrobiology Center, 2-21-1 Osawa, Mitaka, Tokyo 181-8588, Japan}
\altaffiltext{4}{Instituto de Astrof\'{i}sica de Canarias (IAC), 38205 La Laguna, Tenerife, Spain}
\altaffiltext{5}{Department of Earth and Planetary Sciences, Institute of Science Tokyo, 2-12-1 Ookayama, Meguro, Tokyo 152-8551, Japan}
\altaffiltext{6}{Okayama Observatory, Kyoto University, 3037-5 Honjo, Kamogata, Asakuchi, Okayama 719-0232, Japan}
\altaffiltext{7}{National Astronomical Observatory of Japan, 2-21-1 Osawa, Mitaka, Tokyo 181-8588, Japan}
\altaffiltext{8}{Astronomical Science Program, Graduate University for Advanced Studies, SOKENDAI, 2-21-1, Osawa, Mitaka, Tokyo, 181-8588, Japan}
\altaffiltext{9}{Institute for Astronomy, University of Hawai'i, 2680 Woodlawn Drive, Honolulu, HI 96822, USA}
\altaffiltext{10}{Okayama Branch, Subaru Telescope, National Astronomical Observatory of Japan, NINS, Kamogata, Asakuchi, Okayama 719-0232, Japan}
\altaffiltext{11}{Department of Physics and Astronomy, Vanderbilt University, Nashville, TN 37235, USA}
\altaffiltext{12}{Grand Pra Observatory, 1984 Les Hauderes, Switzerland}
\altaffiltext{13}{Wild Boar Remote Observatory, San Casciano in val di Pesa, Firenze, 50026 Italy}
\altaffiltext{14}{INAF - Osservatorio Astrofisico di Catania, Via S. Sofia 78, 95123 Catania, Italy}
\altaffiltext{15}{Privat Observatory Herges-Hallenberg, Steinbach-Hallenberg, Germany}
\altaffiltext{16}{NASA Ames Research Center, Moffett Field, CA 94035, USA}
\altaffiltext{17}{SETI Institute, Mountain View, CA 94043, USA}
\altaffiltext{18}{Headquarter for Co-Creation Strategy, National Institutes of Natural Sciences, Tokyo, 105-0001, Japan}
\altaffiltext{19}{Department of Astronomy, University of Tokyo, 7-3-1 Hongo, Bunkyo, Tokyo 113-0033, Japan}
\footnotetext[$\dag$]{Present address: ....}



\maketitle

\begin{abstract}
Hot Jupiters orbiting hot stars ($T_\mathrm{eff} > 7000$ K) are suggested to have experienced high-eccentricity migration, often evidenced by the tendency for misaligned orbits, despite their circular orbits. In this paper, we present the discovery of TOI-1355 b: an eccentric ($e\sim0.22$) hot Jupiter with a mass of $m_{\mathrm{p}}\sim5.8M_J$ and a radius of $R_{\mathrm{p}}\sim 1.4R_J$ orbiting an A-type star with a period of about $2.17$ days, identified from the TESS transit survey and subsequent follow-up observations. We measured the stellar parameters using the data from the high-resolution spectrograph Seimei/GAOES-RV and obtained the planetary parameters from the photometric data acquired by TESS and ground-based telescopes. This is one of the rare eccentric hot Jupiters around hot stars. This system could be undergoing high-eccentricity migration. We detected nodal precession by measuring the change in its impact parameter. This implies that its transit will no longer be observable from the middle of 2033. Nevertheless, TOI-1355 b is anticipated to be a compelling target for future atmospheric observations, given the hint of atmospheric variability detected in this study. 
\end{abstract}


\section{Introduction}
Hot Jupiters are gas giant planets orbiting close to their host stars ($a/R_{\mathrm{s}} < 10$), which are thought to be difficult to form in situ. Therefore, several orbital evolution mechanisms, such as planet-disk migration and high-eccentricity migration, have been proposed. The former model involves a gas giant being brought close to its host star through gravitational interaction between the planet and the disk, thereby aligning its orbit with the stellar spin (e.g. \cite{1996Natur.380..606L}). The latter describes a model in which a gas giant, whose eccentricity and inclination have been increased by interactions with other planets (planet-planet scattering: e.g. \cite{2008ApJ...686..580C}) or an outer companion star (Kozai-Lidov migration: e.g. \cite{2007ApJ...669.1298F}), approaches its host star because of tidal circularization.

\begin{figure*}[htbp]
 \begin{center}
  \includegraphics[width=\linewidth]{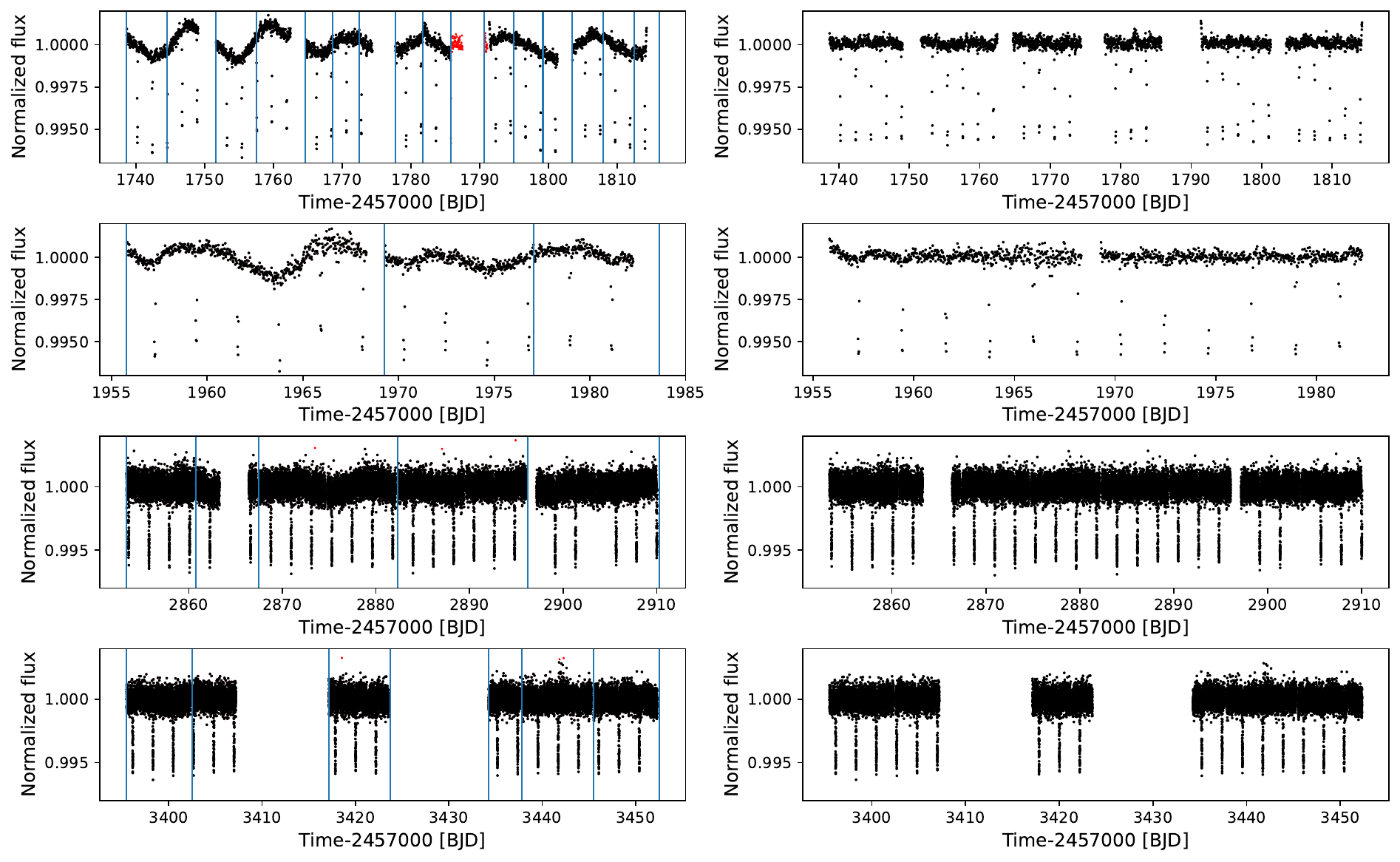}
 \end{center}
\caption{Left: Normalized TESS light curves of TOI-1355 b. From top to bottom, the rows show data from 2019, 2020, 2022 and 2024, respectively. The vertical blue lines show the momentum dumps. The red points are the data points excluded from our analysis due to factors such as small discontinuities, flux ramps, and non-zero quality flag. Right: Light curves after removing the signal attributed to high-amplitude frequencies. {Alt text: 8 light curve graphs aligned with 4 rows and 2 columns. The x axes in all graphs show time in Barycentric Julian Date. The y axes in all graphs show normalized flux.}}\label{TESS_data}
\end{figure*}

\begin{figure*}[htbp]
 \begin{center}
  \includegraphics[width=\linewidth]{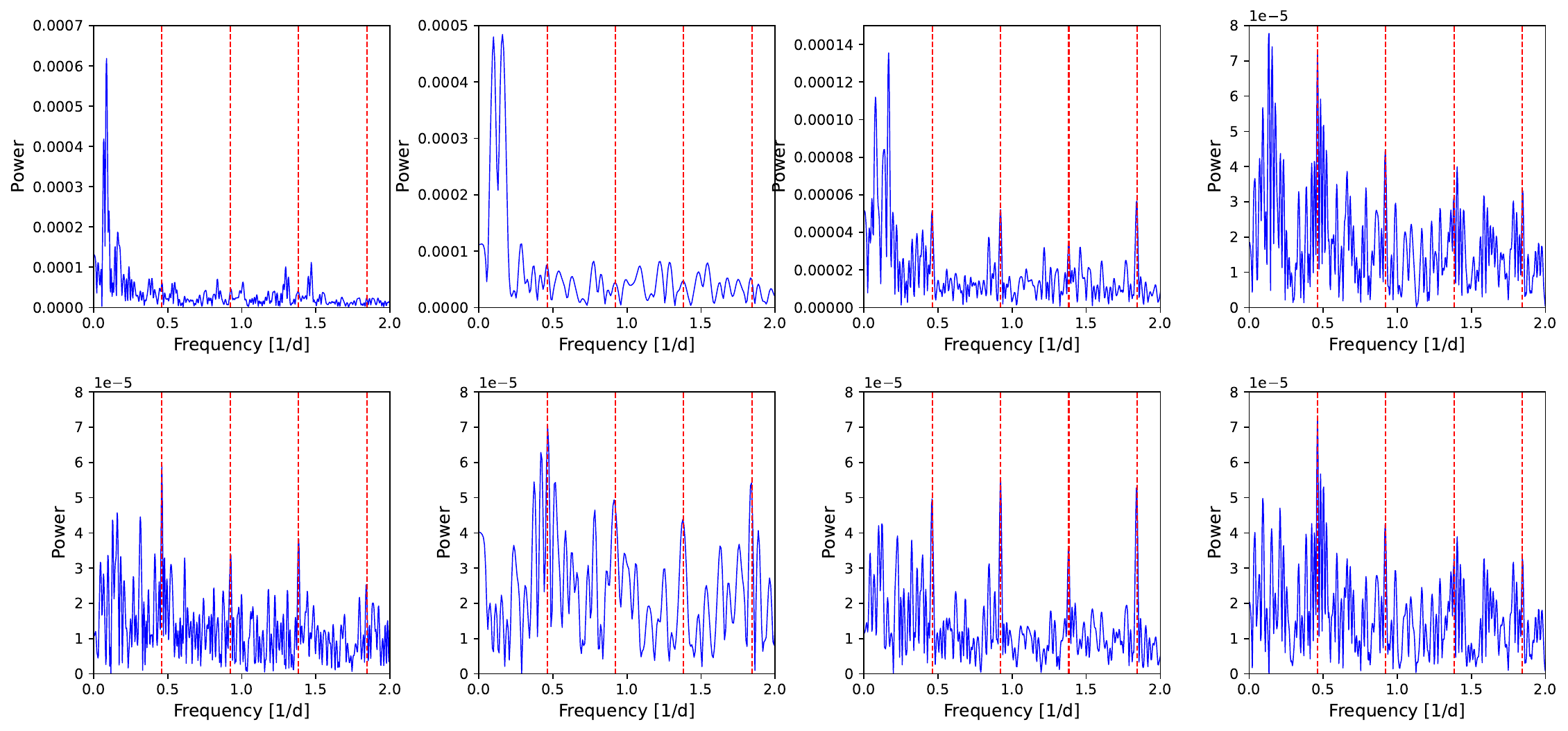}
 \end{center}
\caption{Top: Periodograms for the light curves of TOI-1355 without transits and secondary eclipses parts in 2019 (leftmost), in 2020 (second left), in 2022 (third left), and in 2024 (rightmost). The red dashed lines show the first four harmonics of the orbital period. Bottom: Same as the top panels but for the light curves after removing the high-amplitude frequency components. Signals corresponding to the phase variations are visible. {Alt text: 8 light curve graphs aligned with 2 rows and 4 columns. The x axes in all graphs show frequency in the reciprocal of day. The y axes in all graphs show power.}}\label{TESS_data_per}
\end{figure*}

Because hot stars ($T_\mathrm{eff}>7000$K) lack convective zones in their envelopes and rarely undergo realignment \citep{2012ApJ...757...18A}, systems consisting of a hot Jupiter and a hot star are ideal for investigating orbital evolution. Thus far, about 20 hot Jupiters around hot stars have been discovered by transit surveys such as WASP (Wide Angle Search for Planets; \cite{2006PASP..118.1407P}), KELT (Kilodegree Extremely Little Telescope; \cite{Pepper_2007}) and TESS (Transiting Exoplanet Survey Satellite; \cite{2015JATIS...1a4003R}). These hot Jupiters tend to exhibit misaligned orbits and could have undergone high-eccentricity migration. However, eccentric hot Jupiters ($e>0.1$) around hot stars remain extremely rare, with only a single example (TOI-159 b) discovered recently ($e=0.24\pm0.04$: \cite{2026arXiv260504149M}). The eccentricities of the other three systems of hot Jupiters and hot stars have been measured -- namely MASCARA-1b ($e=0.00034^{+0.00034}_{-0.00023}$: \cite{2022A&A...658A..75H}), Kepler-13Ab ($e=0.00064^{+0.00012}_{-0.00016}$: \cite{2015ApJ...804..150E}) and TOI-1431b ($e=0.0022^{+0.0030}_{-0.0016}$: \cite{2021AJ....162..292A}) -- and all of these values are close to $e=0$. The eccentricities of other such systems are assumed to be zero.
The discovery of hot Jupiters that retain eccentric orbits before undergoing circularization is essential for validating the high-eccentricity migration pathway around hot stars.

Our paper presents the discovery of TOI-1355 b, an eccentric hot super-Jupiter orbiting an A-type star. This paper is structured as follows. Section 2 presents observations of TOI-1355 b using photometric instruments (TESS, MuSCAT3, RCO 40 cm, OACT 91 cm, and SCT in Herges-Hallenberg), a high-resolution spectrograph (GAOES-RV), and a speckle imaging technique (NESSI). Section 3 describes the analysis of the stellar characterization using spectral data. Section 4 explains the methodology for measuring the parameters of the TOI-1355 system from light curve fitting to photometric data. Sections 5 and 6 present the results and discussion on TOI-1355 b, respectively. Finally, we present our conclusions in section 7.

\section{Observations}
\subsection{TESS Photometry}
TESS observed TIC 372264750 during the period of UT 2019 September 12 -- November 27 (Sectors 16, 17 and 18), UT 2020 April 16 -- May 12 (Sector 24), UT 2022 September 30 -- November 26 (Sectors 57 and 58), and UT 2024 March 25 -- May 21 (Sectors 77 and 78). This system was designated as a TESS Object of Interest (TOI) with a number of 1355. We opted to use TESS-SPOC (Science Processing Operations Center) HLSP (High-Level Science Products; \cite{2020RNAAS...4..201C}) light curves of TOI-1355 in 2019 and 2020 since 2-minute cadence SPOC light curves were unavailable for Sectors 16, 17, 18 and 24.
On the other hand, we utilized SPOC light curves \citep{jenkinsSPOC2016} of TOI-1355 in 2022 and 2024. These datasets were acquired from the Mikulski Archive for Space Telescopes (MAST) via \texttt{lightkurve} \citep{2018ascl.soft12013L}. The exposure times were 1800 s for the 2019 and 2020 data, and 120 s for the 2022 and 2024 data. These data include the Presearch Data Conditioning SAP\footnote{SAP is an abbreviation for Simple Aperture Photometry.} (PDCSAP) light curve (\cite{Stumpe2012}; \cite{Stumpe2014}), which is corrected for systematic artifacts. We excluded portions of light curves that exhibited small discontinuities and flux ramps caused by the momentum dumps. The left panel of figure \ref{TESS_data} shows the light curves of TOI-1355 obtained from the TESS data.

The photometric aperture used by PDCSAP is contaminated by the light from TIC 372264743 (also known as ZTF J220522.25+664644.0), an RS Canum Venaticorum variable star with a period of $\sim 11.56$ days \citep{2020ApJS..249...18C}, which distorts the PDCSAP light curve of TOI-1355. We first created periodograms of TOI-1355 light curves, excluding transits and secondary eclipses. 
We then iterated the process to remove the highest-amplitude frequency using the function \texttt{to\_periodogram} included in \texttt{lightkurve} \citep{2018ascl.soft12013L}. We subtracted long-term components that occurred on timescales longer and with amplitudes larger than those associated with the orbital period. For the frequency removal process, a total of eight, five, four and one dominant components were removed from the 2019, 2020, 2022 and 2024 data, respectively.
The right panel of figure \ref{TESS_data} shows the flattened light curve, and figure \ref{TESS_data_per} displays the periodograms of the TESS data.

\subsection{Ground-based photometry}
\subsubsection{MuSCAT3}
We observed a full transit of TOI-1355 b simultaneously in the $g$ (400-550 nm), $r$ (550-700 nm), $i$ (700-820 nm) and $z_{s}$ (820-920 nm) bands using MuSCAT3 \citep{2020SPIE11447E..5KN} mounted on the 2 m Faulkes Telescope North at Haleakale Observatory, Maui, Hawaii (USA). MuSCAT3 has a 9.1$'$ $\times$ 9.1$'$ field of view. These data were obtained on 2024 July 4. The exposure time of $g$, $r$, $i$ and $z_{s}$ bands was 3, 3, 2 and 2.6 s, respectively.

The raw images were reduced with \texttt{BANZAI} pipeline \citep{2018SPIE10707E..0KM}. We then performed aperture photometry using the custom pipeline \citep{2011PASJ...63..287F}. HD210166, the brightest star in the frame, was chosen as a comparison star to remove atmospheric effects. We adopted aperture radius of 32 pixels for the $g$ and $r$ bands, and 34 pixels for the $i$ and $z_{s}$ bands. 

\subsubsection{RCO 40cm}
Two transit observations of TOI-1355 b were obtained on 2019 November 6 and 2019 November 19 with the FLI 4710 camera mounted on an RCO 40 cm telescope located at the Grand-Pra Observatory (Switzerland). FLI 4710 is an 11.7$'$ $\times$ 11.7$'$ field of view back-illuminated CCD using an E2V CCD47-10 sensor. Observations were taken in 1 $\times$ 1 binned mode and produced a 0.730$''$ pixel scale. We observed a full transit in the Sloan $z_s$ passband and a transit ingress in the Sloan $g'$ passband both with an exposure time of 60 s.  We produced the light curve of TOI-1355 b using the AstroImageJ (AIJ) software with 6.2$''$ and 10.0$''$ apertures, respectively.

\subsubsection{OACT 91cm}
TOI-1355 b photometric data were obtained on 2019 November 30 with the Cassegrain 91 cm telescope of "M. G. Fracastoro" observing station of Catania Astrophysical Observatory (INAF-OACT), situated at 1750 m on Mt. Etna (Italy). The CCD used was a KAF 1001E (24 $\mu$m $\times$ 24 $\mu$m - 1024 $\times$ 1024 pixels). The field of view is approximately 11$'$ $\times$ 11$'$. We used $B$ Johnson-Bessel filter, and set the exposure time to 60 s. The read-out time was 10 s, and the mean FWHM of the target star was 12$''$.

Data reduction was performed via AIJ software, selecting an optimized aperture radius of 24 pixels (=16$''$) and using the unique suitable comparison star present in the small field of view. 

\subsubsection{SCT 280/1870 in Herges-Hallenberg}
The photometric observation for TOI-1355 b's transit took place on 2022 November 19 at a private observatory in Herges-Hallenberg (Germany). The images were taken with a Moravian G2-1600 on an SCT 280/1870 (Celestron C11 with Starizona reducer). The exposure time was 90 s using an $I$ filter. The images were corrected with dark and flat. AIJ was used for the photometric analysis.

\subsection{GAOES-RV Spectroscopy}
We obtained spectral data of TOI-1355 on 2023 July 27 and 2023 December 13, during TOI-1355 b's transit, using GAOES-RV \citep{2024SPIE13096E..44S} installed on the 3.8-m Seimei telescope in Okayama (Japan) \citep{2020PASJ...72...48K}. This spectrograph provides a resolution of $R\sim 65,000$ and a wavelength range of $516 - 593$ nm. The datasets in July and December contain 12 and 15 spectra, respectively. The exposure time for each spectrum was 1200 s. The signal-to-noise ratio (SNR) per pixel at 550 nm for each spectrum was $\sim80$ and $\sim60$ for the July and December datasets, respectively.
The datasets were reduced to 1D spectra during observation\footnote{http://www.o.kwasan.kyoto-u.ac.jp/inst/gaoes-rv/dataanalysis.html}. A wavelength range between 517.5 nm and 566.5 nm was utilized, excluding the range of low SNR and bad pixels. We then performed normalization of the spectral datasets and the removal of the telluric lines using \texttt{telfit} \citep{2014AJ....148...53G}. Finally, we combined the out-of-transit spectral datasets in the two epochs. It is noted that the first exposure of the July spectral dataset was not used for combining due to its low SNR ($\sim 25$).

We performed a Doppler tomographic analysis using these data. However, the SNRs in both epochs were not high enough to detect a planetary signature. Thus, we used the out-of-transit datasets to measure the stellar parameters.

\subsection{NESSI Speckle Imaging}
\begin{figure}[htbp]
 \begin{center}
  \includegraphics[width=\linewidth]{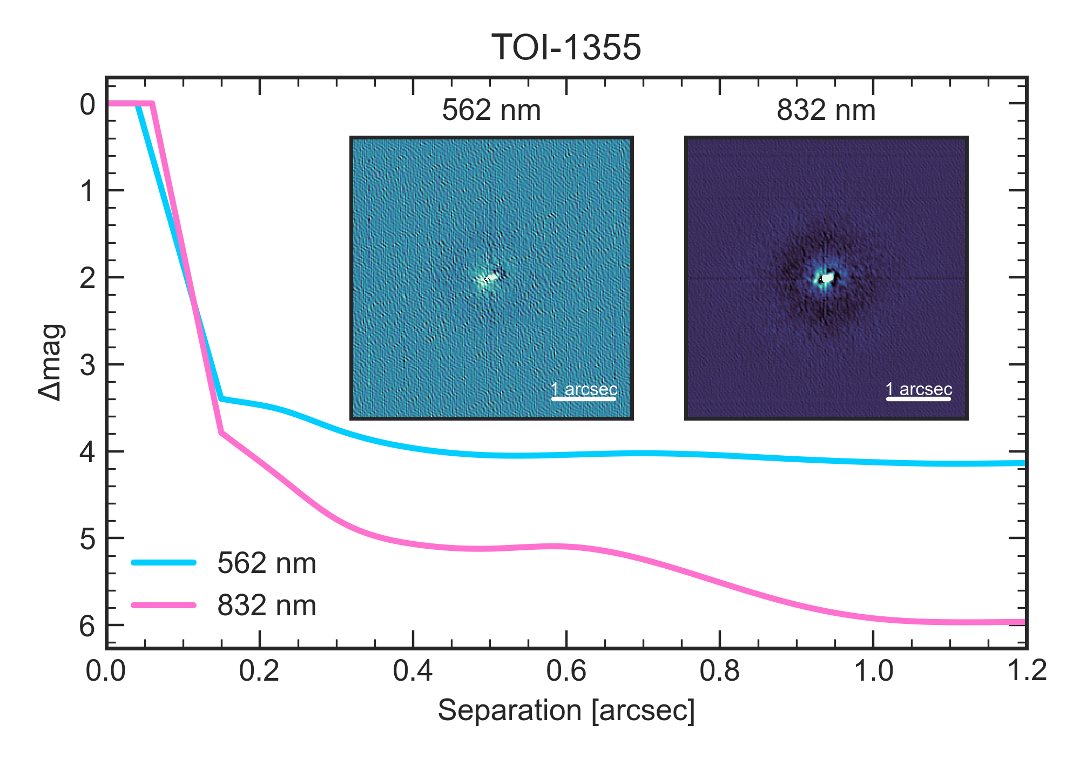}
 \end{center}
\caption{5σ contrast curves from NESSI speckle imaging for TOI-1355. The upper cyan line and lower magenta line are the blue band centered at 562 nm and the red band centered at 832 nm, respectively. The insets display 4.6 \arcsec $\times$ 4.6 \arcsec images. No companions are detected in these images. {Alt text: There is one graph. The x axis shows the separation. The y axis shows the magnitude difference. There are 2 images in the graph. Each image shows the target in the center.}}
\label{fig:wiyn}
\end{figure}

On the night of 2019 November 9 UT, TOI-1355 was observed with the NESSI speckle imager \citep{Scott2019}, mounted on the 3.5\,m WIYN telescope at Kitt Peak. NESSI simultaneously acquires data in two bands centered at 562 nm and 832 nm using high speed electron-multiplying CCDs (EMCCDs). We collected and reduced the data following the procedures described in \citet{Howell2011}. The resulting reconstructed image achieved a contrast of $\Delta\mathrm{mag} \sim 6$ at a separation of 1\arcsec\ in the 832 nm band (see figure~\ref{fig:wiyn}). 

\section{Stellar Characterization}
\subsection{SME Analysis}
\begin{figure}[htbp]
 \begin{center}
  \includegraphics[width=\linewidth]{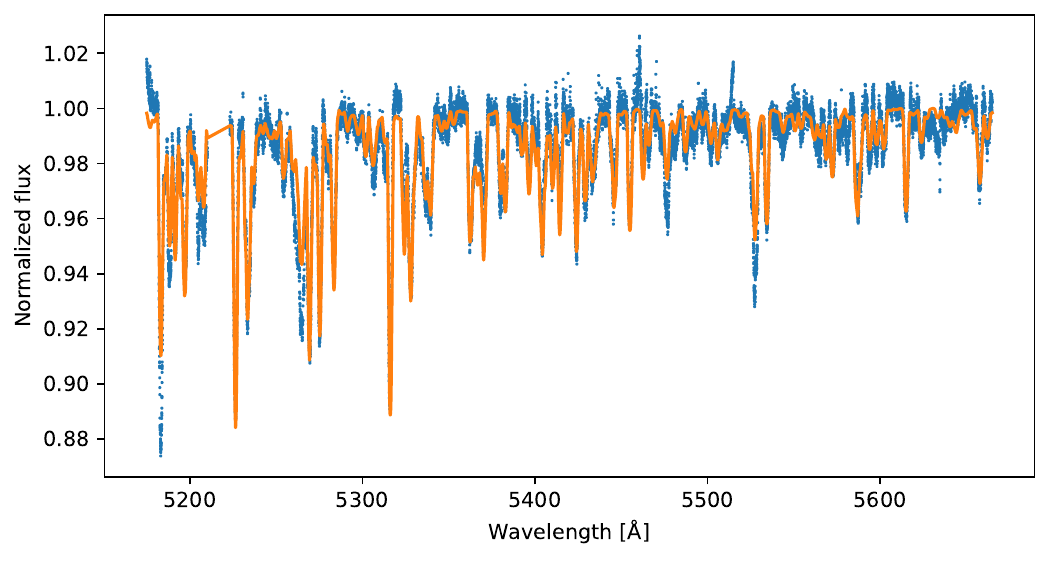}
 \end{center}
\caption{The spectral lines of TOI-1355 are shown with blue dots representing the observed spectral data from Seimei/GAOES-RV, and the orange line denoting the spectral model derived from SME. {Alt text: There is one graph. The x axis shows the wavelength. The y axis shows relative flux.}}\label{Spec}
\end{figure}

We first inferred the stellar parameters using the spectral datasets in subsection 2.3. This inference was conducted with \texttt{PySME} \citep{2023A&A...671A.171W}, a tool based on \texttt{Spectroscopy Made Easy} (SME; \cite{1996A&AS..118..595V}, \cite{2017A&A...597A..16P}). The model was constructed by combining VALD3 line list \citep{2017ASPC..510..518P}, an \texttt{ATLAS9} atmospheric model \citep{2002A&A...392..619H, 2017ascl.soft10017K}, and a Gaussian instrumental profile with a resolution of $R\sim 65,000$.
Figure \ref{Spec} displays the best-fit spectral model derived from SME.
In this analysis, we derived stellar effective temperature ($T_{\rm eff} = 8780 \pm 350$ K), surface gravity ($\log g = 3.81 \pm 0.52$) overall metallicity ([M/H]$= 0.03 \pm 0.17$), projected rotational velocity ($v\sin i_{\rm s}= 80.8 \pm 2.7$ km s$^{-1}$), micro- and macro-turbulent velocity ($\xi_{\rm mic}= 2.87 \pm 0.51$ km s$^{-1}$, $\xi_{\rm mac}= 22.9 \pm 5.4$ km s$^{-1}$). For the calculation of [M/H], we adopted the solar abundances from \citet{1998SSRv...85..161G} as the reference, assuming a scaled solar abundance pattern.
Note that \texttt{PySME} could not quantify the uncertainty of the radial velocity. However, its value was determined to be $v_{\mathrm{rad}}=-25.56$ km s$^{-1}$. 

\subsection{SED Analysis by ARIADNE}
We then performed spectral energy distribution (SED) fitting using \texttt{ARIADNE} Python package \citep{2022MNRAS.513.2719V} to investigate additional stellar parameters: stellar radius, mass, age, and distance to the system. Specifically, stellar radius and mass are essential for measuring the planetary radius and mass. We used Kurucz stellar atmosphere models \citep{1993sssp.book.....K} to model the SED. This was done using $B$ and $V$ magnitudes from Tycho-2, W1 and W2 magnitudes from WISE, and the $G$, $G_{BP}$ and $G_{RP}$ magnitudes from Gaia. From the \texttt{ARIADNE} pipeline, stellar effective temperature, surface gravity, metallicity [Fe/H], extinction ($A_v$), stellar radius, mass, age, and distance from the Solar system were calculated. In this analysis, we used the solar abundances from \citet{1998SSRv...85..161G} as the reference for [Fe/H], and assumed a scaled solar abundance pattern. Note that we adopted the results for the stellar effective temperature and surface gravity from section 3.1 as their respective Gaussian priors, while the Gaussian prior for [Fe/H] was informed by the result obtained for [M/H] in the same section.

\begin{figure}[htbp]
 \begin{center}
  \includegraphics[width=\linewidth]{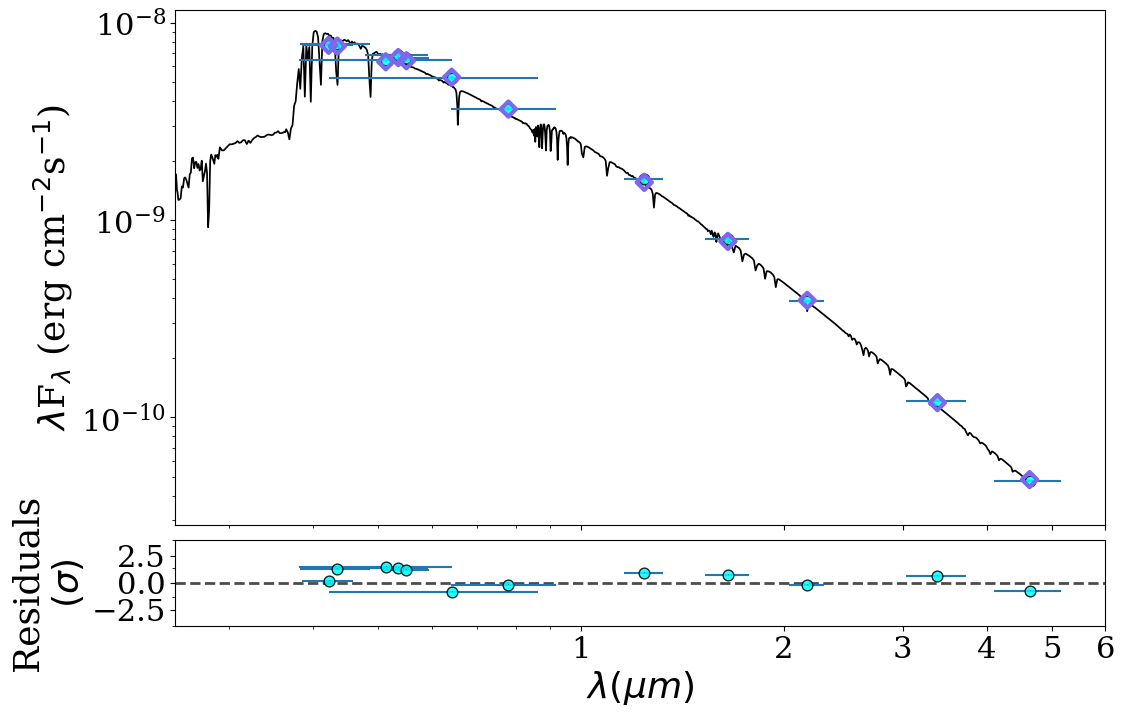}
 \end{center}
\caption{Top: Spectral energy distribution of the target star TOI-1355 with, magnitudes from the catalogs listed in table \ref{stellar_tab} displayed as blue dots. Black line represents the spectrum generated from Kurucz stellar atmosphere models. Bottom: The difference between the fluxes (derived from magnitudes) and the spectral model. {Alt text: There are two graphs. The x axes of the 2 graphs shows the wavelength. The y axis in the top graph shows flux. The y axis in the bottom graph shows residuals scaled by standard deviation.}}\label{SED}
\end{figure}

The properties of TOI-1355, magnitudes from the photometric catalog, and derived stellar values are listed in table \ref{stellar_tab}, along with their uncertainties, and figure \ref{SED} shows the best-fitting SED model. We calculated the stellar effective temperature ($T_{\rm eff} = 8660^{+360}_{-310}$ K), surface gravity ($\log g = 3.95^{+0.44}_{-0.53}$), iron abundance ([Fe/H]$= -0.01^{+0.17}_{-0.19}$), distance ($D=246.98^{+0.96}_{-1.00}$ pc), stellar radius ($R_{\mathrm{s}}=1.881^{+0.048}_{-0.056}\ R_{\odot}$), stellar mass ($M_{\mathrm{s}}=2.00^{+0.21}_{-0.13}\ M_{\odot}$), stellar age ($0.37^{+0.21}_{-0.36}$ Gyrs), and extinction ($A_v=0.163^{+0.089}_{-0.099}$ mag).

\begin{table}[htbp]
  \tbl{Stellar parameters of TOI1355.}{
  \begin{tabular}{lccc}
      \hline
      Identifier &&&  \\ 
      \hline
      TIC & 372264750&&\\
      HD & 210058&&\\
      2MASS & J22051380+6646308&&\\
      \hline
      Coordinate &&&  \\ 
      \hline
      R.A. (J2000)& \timeform{22h05m13s.79}&&\\
      Dec (J2000) & \timeform{+66D46'30''.95}&&\\
      \hline
      Magnitude &&&  \\ 
      \hline
      TYCHO B $B_T$   &  $8.893\pm0.016$  &&\\
      TYCHO V $V_T$   &  $8.721 \pm 0.013$ &&  \\
      Gaia BP $G_\mathrm{BP}$ & $8.7616\pm0.0029$  && \\
	  Gaia G $G$ &	 $8.7152 \pm 0.0028$ &&  \\
	  Gaia RP $G_\mathrm{RP}$    &	 $8.6077 \pm 0.0038$ &&  \\
	  2MASS J     & $8.444 \pm 0.024$ && \\
	  2MASS H     & $8.424 \pm 0.033$ && \\
	  2MASS Ks    &	$8.442 \pm 0.024$ && \\
      WISE RSR W1 & $8.410 \pm 0.023$ && \\
      WISE RSR W2 & $8.438 \pm 0.019$ && \\
      \hline
      Stellar Parameter & \texttt{ARIADNE} & \texttt{PySME} & Prior for \texttt{ARIADNE}\\ 
      \hline
      $T_{\mathrm{eff}}$ (K)& $8660^{+360}_{-310}$&$8780\pm350$&$\mathcal{N}(8780,350)$\\
      $\log g$ (cm s$^{-2}$)  & $3.95^{+0.44}_{-0.53}$&$3.81\pm0.52$ & $\mathcal{N}(3.81,0.52)$\\
      \ [M/H] &-&$0.03\pm0.17$&-\\
      \ [Fe/H] & $-0.01^{+0.17}_{-0.19}$&-& $\mathcal{N}(0.03,0.17)$\footnotemark[$*$]\\
      $D$ (pc) &$246.98^{+0.96}_{-1.00}$&-&$\mathcal{N}(246.85,0.92)$\footnotemark[$\dag$] \\
      $R_{\mathrm{s}}$ ($R_\odot$)& $1.881^{+0.048}_{-0.056}$&-& $\mathcal{U}(0.05,100)$\\
      $M_{\mathrm{s}}$ ($M_\odot$)\footnotemark[$\S$]&  $2.00^{+0.21}_{-0.13}$&-&- \\
      Age (Gyrs)\footnotemark[$\S$]&  $0.37^{+0.21}_{-0.36}$&-&-\\
      $A_v$ (mag)&  $0.163^{+0.089}_{-0.099}$ &-&$\mathcal{U}(0,1.814)$\\
      $v\sin i_{\rm s}$ (km s$^{-1}$)&-&  $80.8\pm2.7$&- \\
      $\xi_\mathrm{mic}$ (km s$^{-1}$)&-&  $2.87\pm0.51$&- \\
      $\xi_\mathrm{mac}$ (km s$^{-1}$)&-&  $22.9\pm5.4$&- \\
      Spectral Type &A3V&-&-\\
      \hline
    \end{tabular}}
\label{stellar_tab}
\begin{tabnote}
\footnotemark[$\S$] These values are calculated from isochrone analysis via \texttt{ARIADNE}. \\
\footnotemark[$*$] Means and standard deviations of [M/H] from \texttt{PySME} are adopted for the Gaussian prior of [Fe/H]. \\
\footnotemark[$\dag$] Means and standard deviations are calculated from Gaia Parallax (4.051 $\pm$ 0.015 mas).
\end{tabnote}
\end{table}

\section{Photometric Light Curve Analysis}
\subsection{Phase Curve Model of Full-orbit}
The TESS light curves, which remove the component of the nearby variable stars, represent the variability of the TOI-1355 system, including primary transits, secondary eclipses, and variations associated with the orbital phase. The full-orbit model, the total flux of the entire system $F_{\mathrm{full}}/F_{\mathrm{s}}$, is expressed as 
\begin{equation}
\label{LC}
\frac{F_{\mathrm{full}}}{F_{\mathrm{s}}} = \left\{f_{\mathrm{TR}}\left(1 + \frac{F_{\mathrm{DB}}}{F_{\mathrm{s}}} + \frac{F_{\mathrm{EV}}}{F_{\mathrm{s}}}\right) + \varepsilon \left(\frac{F_{\mathrm{TE}}}{F_{\mathrm{s}}}+\frac{F_{\mathrm{RE}}}{F_{\mathrm{s}}}\right) \right\}B,
\end{equation}
with $B$ as a constant. Here, $f_{\mathrm{TR}}$ is the transit function defined by subtracting the fraction of flux lost during transit from an out-of-transit flux normalized to unity; $F_{\mathrm{DB}}$ refers to Doppler boosting; $F_{\mathrm{EV}}$ denotes ellipsoidal variations of the host star; $F_{\mathrm{TE}}$ represents thermal emission from the planet; $F_{\mathrm{RE}}$ indicates reflection from the planet, and $\varepsilon$ is the eclipse function scaled from 0 (full eclipse) to 1 (out of eclipse). $f_{\mathrm{TR}}$ is modeled by \texttt{PyTransit} \citep{Parviainen2015} using the ratio of the planetary radius $R_{\mathrm{p}}$ to stellar radius $R_{\mathrm{s}}$, quadratic limb-darkening coefficients $u_1$ and $u_2$, transit epoch $T_c$, orbital period $P_{\mathrm{orb}}$, ratio of semi-major axis $a$ to $R_{\mathrm{s}}$, orbital inclination $i_{\mathrm{p}}$, eccentricity $e$, and argument of periastron $\omega$. Note that $i_{\mathrm{p}}$ can be expressed in terms of the impact parameter $b$ using
\begin{equation}
\label{angle_ip}
\cos i_{\mathrm{p}}=\frac{b R_{\mathrm{s}}}{a}\frac{1+e\sin\omega}{1-e^2}.
\end{equation}

\subsubsection{Reflection}
As described by \citet{2022A&A...668A..93P}, $F_{\mathrm{RE}}$ is given by the following equation:
\begin{equation}
\label{Day_1}
\frac{F_{\mathrm{RE}}}{F_{\mathrm{s}}}=A_{\mathrm{g}}\left(\frac{R_{\mathrm{p}}}{R_{\mathrm{s}}}\frac{R_{\mathrm{s}}}{a}\right)^2\left(\frac{1+e\cos{f}}{1-e^2}\right)^2\Phi,
\end{equation}
where $A_{\mathrm{g}}$ and $f$ are the geometric albedo and true anomaly, respectively.
We assumed the phase-function for the thermal emission $\Phi$ as a Lambertian, given by
\begin{equation}
\label{Phi}
\Phi=\frac{\sin \alpha -\alpha\cos{\alpha}}{\pi},
\end{equation}
where $\alpha$ is the phase angle. $\alpha$ can be described as
\begin{equation}
\label{angle_dash}
\cos\alpha=\sin(f+\omega)\sin{i_{\mathrm{p}}}.
\end{equation}

\subsubsection{Thermal Emission}
Thermal emission from a planet can be modeled as a sine function of $\alpha'$, which is described by 
\begin{equation}
\label{angle}
\cos\alpha'=\sin(f+\omega+\delta)\sin{i_{\mathrm{p}}}.
\end{equation}
This $\alpha'$ is different from $\alpha$ due to the influence of the planetary phase offset $\delta$, which sets the location of the hot spot.
Thus, $F_{\mathrm{TE}}/F_{\mathrm{s}}$ can be expressed as 
\begin{equation}
\label{PE}
\frac{F_{\mathrm{TE}}}{F_{\mathrm{s}}}=\frac{F_{N}}{F_{\mathrm{s}}}+\frac{1}{2}\left(\frac{F_{D}}{F_{\mathrm{s}}}-\frac{F_{N}}{F_{\mathrm{s}}}\right)(1-\cos\alpha').
\end{equation}
including night-side thermal emission $F_{N}$, day-side thermal emission $F_{D}$.

Treating $F_{N}$, $F_{D}$ and $F_{\mathrm{s}}$ as black body radiation, we can write $F_{N}/F_{\mathrm{s}}$ and $F_{D}/F_{\mathrm{s}}$ as follows,
\begin{align}
\label{BB}
\frac{F_{N}}{F_{\mathrm{s}}}&=\frac{\int \tau (\lambda) F_{B}(T_{N}, \lambda)d\lambda}{\int \tau (\lambda) F_{B}(T_{\mathrm{eff}}, \lambda)d\lambda}\left(\frac{R_{\mathrm{p}}}{R_{\mathrm{s}}}\right)^2.\\
\frac{F_{D}}{F_{\mathrm{s}}}&=\frac{\int \tau (\lambda) F_{B}(T_{D}, \lambda)d\lambda}{\int \tau (\lambda) F_{B}(T_{\mathrm{eff}}, \lambda)d\lambda}\left(\frac{R_{\mathrm{p}}}{R_{\mathrm{s}}}\right)^2.
\end{align}
In these equations, $\tau (\lambda)$ represents the TESS passband transmission, $F_{B}$ denotes the flux spectrum determined by the Planck's law, $T_{N}$ is the night-side planetary brightness temperature, $T_{D}$ is the day-side planetary brightness temperature, and $\lambda$ is the wavelength.
Assuming $T_N$ and $T_D$ are inversely proportional to the square root of distance between the planet and host star as in \citet{2011ApJ...729...54C}, they are given by the following equations:
\begin{align}
\label{TNTD}
T_N&=T_{N, a} \sqrt{\frac{1+e\cos f}{1-e^2}}\\
T_D&=T_{D, a} \sqrt{\frac{1+e\cos f}{1-e^2}},
\end{align}
where $T_{N, a}$ and $T_{D, a}$ are the day-side and night-side planetary brightness temperatures at a distance of $a$ from the host star.

\subsubsection{Doppler Boosting}
The semi-amplitude of the radial velocity $K$ is described by \citet{1999ApJ...526..890C} as
\begin{align}
\label{DB_1}
K=\left(\frac{2\pi G}{P}\right)^{1/3}\frac{m_{\mathrm{p}}\sin i_{\mathrm{p}}}{M_{\mathrm{s}}^{2/3}}\frac{1}{\sqrt{1-e^2}}.
\end{align}
Therefore, the equation for the phase curve due to Doppler boosting is 
\begin{align}
\label{DB_2}
\frac{F_{\mathrm{DB}}}{F_{\mathrm{s}}}&=-\frac{\beta}{c}K\{\cos (f+\omega)+e\cos (\omega)\} \nonumber \\
&=-\frac{\beta}{c}\left(\frac{2\pi G}{P}\right)^{1/3}\frac{m_{\mathrm{p}}\sin i_{\mathrm{p}}}{M_{\mathrm{s}}^{2/3}}\frac{1}{\sqrt{1-e^2}}\{\cos (f+\omega)+e\cos (\omega)\},
\end{align}
where $c$ is the speed of light, $m_{\mathrm{p}}$ is the planetary mass, and $G$ is the gravitational constant. Following \citet{2011MNRAS.410.1787B} and \citet{2003ApJ...588L.117L}, the beaming factor $\beta$ is defined as
\begin{align}
\label{DB_3}
\beta=\frac{\int \tau (\lambda) \lambda F_{\lambda} (5+d\log{F_{\lambda}/d\log{\lambda}) d\lambda}}{\int \tau (\lambda) \lambda F_{\lambda}d\lambda},
\end{align}
where $F_{\lambda}$ is the stellar flux. We calculated $\beta$ using the code from \citet{2021A&A...645A..16P}.

\subsubsection{Ellipsoidal Variation}
Following \citet{1959cbs..book.....K}, we can approximately write the equation of ellipsoidal variation as 
\begin{equation}
\label{EV}
\frac{F_{\mathrm{EV}}}{F_{\mathrm{s}}}=\frac{3}{4}w\frac{m_{\mathrm{p}}}{M_{\mathrm{s}}}\left(\frac{R_{\mathrm{s}}}{a}\frac{1+e\cos f}{1-e^2}\right)^{3}\sin^2 i_{\mathrm{p}} \cos2(f+\omega).
\end{equation}
The coefficient $w$ is expressed as
\begin{equation}
\label{coe_w}
w=\frac{2(15+u_1)(1+g)}{5(6-2u_1-3u_2)},
\end{equation}
where $g$ is the gravity darkening coefficient, set to 0.25 \citep{1924MNRAS..84..665V} in this analysis.

\subsection{Joint Photometric Analysis of TESS and Ground-based Telescopes}
Light curve models for the TESS datasets were generated as described in subsection 4.1. For this fitting, the logarithm of the likelihood for each year's TESS light curve was given by 
\begin{eqnarray}
\label{like_T}
\ln L_{\mathrm{like,T}} =-\frac{1}{2} \sum_{i} \left[\frac{\left(O_{i}-C_{i}\right)^2}{\sigma_i^2+\sigma_{\mathrm{jit}}^2}+ \ln \left\{2 \pi \left(\sigma_i^2+\sigma_{\mathrm{jit}}^2\right)\right\} \right],
\end{eqnarray}
where $i$ denotes the $i$th data point, $O_i$ is the data, $C_i$ is the model, $\sigma_i$ represents the error of the $i$th data point, and $\sigma_{\mathrm{jit}}$ is the jitter.

In addition, transit light curve models with the trends due to the airmass and shifts of x and y were constructed using \texttt{PyTransit} \citep{Parviainen2015}. The logarithm of the likelihood for the light curve of each telescope, $\ln L_{\mathrm{like,G}}$, can be written as in Equation (\ref{like_T}). 
The light curve model with the trends is described with the trend as 
\begin{eqnarray}
\label{trend}
f_\mathrm{TR}(D_B+D_t \Delta t +D_A A_M+D_x \Delta x+D_y \Delta y),
\end{eqnarray}
where $D_B$, $D_t$, $D_A$, $D_x$ and $D_y$ are the constances, $\Delta t$ denotes the time from the beginning of the observation, $A_M$ indicates the airmass, and $\Delta_x$ and $\Delta_y$ are the shifts of $x$ and $y$.

The light curves of TESS and ground-based telescopes were fitted jointly using the Markov Chain Monte Carlo (MCMC) with \texttt{emcee} \citep{2013PASP..125..306F}. 
We performed this analysis under two conditions: the lack of clouds and the existence of clouds on the day-side. For the former case, we hypothesized that the day-side flux consists of only thermal emission ($A_g=0$: hereafter Cloud-free Case). On the other hand, for the latter case, we assumed that the day-side flux contains a reflection component ($A_g>0$: hereafter Cloudy Case).
Across both cases, $T_c$, $b$, $\omega$, $T_{D,a}$, $T_{N,a}/T_{D,a}$, $\delta$ and $B$ of each epoch, $R_{\mathrm{p}}/R_{\mathrm{s}}$, $u_1$ and $u_2$ of each band\footnote{We assume Gaussian priors for $u_1$ and $u_2$. Their means are referred to values calculated by \texttt{PyLDTk} \citep{Husser2013, Parviainen2015_b} using $T_{\mathrm{eff}}$, $\log g$ and [Fe/H] in table \ref{stellar_tab}.}, $\sigma_{\mathrm{jit}}$ for each light curve, $D_B$, $D_t$, $D_A$, $D_x$ and $D_y$ for each light curve of ground-based telescope, $P_{\mathrm{orb}}$, $e$, $m_{\mathrm{p}}$, $a/R_{\mathrm{s}}$, $M_{\mathrm{s}}$ and $T_{\mathrm{eff}}$ were set as free parameters.
Additionally, $A_g$ of each epoch were treated as free parameters for the Cloudy Case.
The orbital parameters $b$ and $\omega$ could vary over several years due to the precession of the planetary orbit by the oblate host star, whereas $e$ and $a$ remain constant \citep{2011PhRvD..84l4001I}.
In these fittings, the logarithm of the probability for the TESS and other telescope datasets was defined as
\begin{align}
\label{post_tr}
\ln L_{\mathrm{prob}} &=\sum_{Y} \ln L_{\mathrm{like,T}}+\sum_{T_e} \ln L_{\mathrm{like,G}}-\frac{1}{2}\sum_{j} \left(\frac{p_{j}-\mu_{j}}{s_j}\right)^2,
\end{align}
where $Y$ and $T_{e}$ indicate the years and the ground-based telescopes, respectively. The third term represents the Gaussian priors: where $p_j$ is the parameter value, and $\mu_j$ and $s_j$ are the center and uncertainty of the Gaussian prior. After MCMC sampling with a total of 500,000 steps per walker using 250 walkers, we discarded the first 200,000 steps as burn-in. We then computed the integrated autocorrelation length, which was approximately 5,000 steps for each parameter, using the iterative procedure described by \citet{Sokal1996MonteCM}, as implemented in the \texttt{emcee} package. The remaining post-burn-in chain length of 300,000 steps exceeded 50 times the autocorrelation lengths for all parameters, following the standard recommendation for the \texttt{emcee} package (indicating that the chains had converged). For the final analysis, the chains were then thinned by a factor of 30, yielding a final length of 10,000 steps per walker.

\section{Results}
\begin{figure*}
\begin{center}
\includegraphics[width=\linewidth]{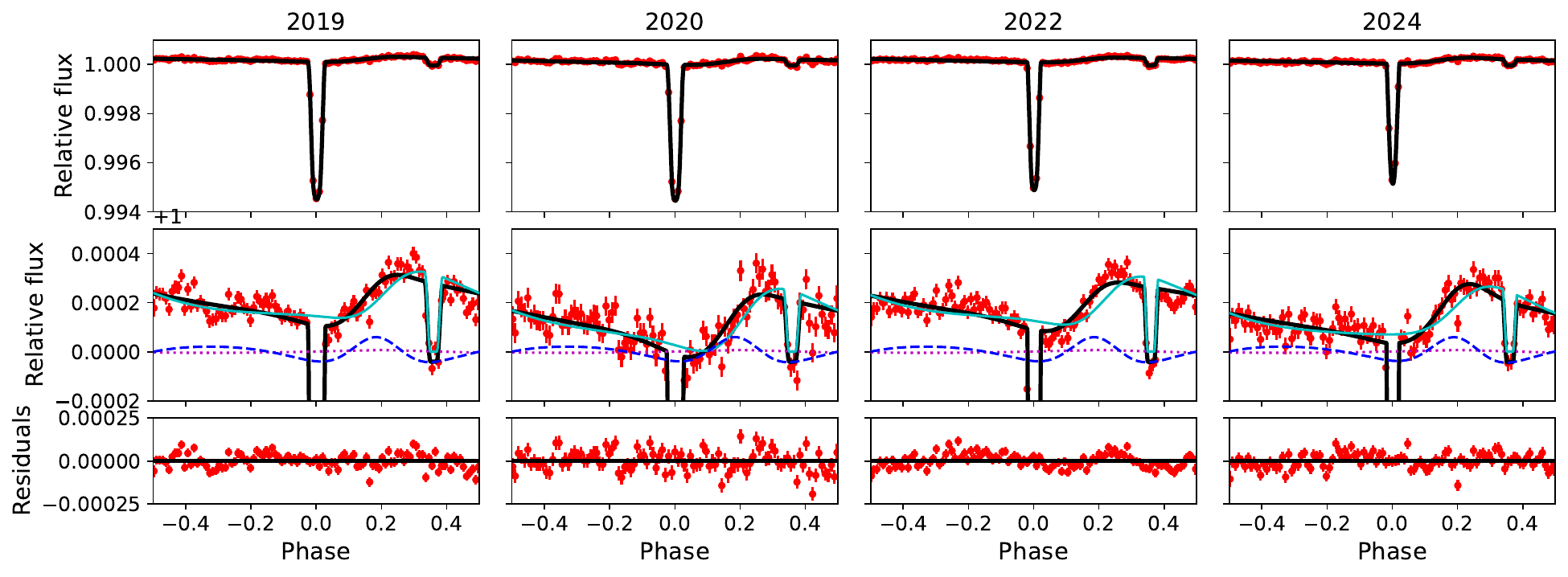}
\end{center}
\caption{Phase-folded light curves of TOI-1355 b from the TESS data in 2019 (leftmost), in 2020 (second left), in 2022 (third left), and in 2024 (rightmost) when Cloud-free Case. Phase, the parameter of x-axis, is normalized by orbital period and its zero is defined as mid-transit time. Top panel: The whole phase curve. Middle panel: The out-of-transit variation. The components of planetary emission, Doppler boosting and ellipsoidal variation are displayed as the thin cyan lines, the magenta dotted lines and the blue dashed lines, respectively. Bottom panel: The residuals between the observed data and model data. The red points show the 30-minute binned data, respectively. The black lines show the best-fit light curve models. {Alt text: 12 graphs aligned with 3 rows and 4 columns. The x axes in all graphs show phase. The y axes of 4 graphs in the top row show relative flux. The y axes of 4 graphs in the middle row show magnified relative flux. The y axes of 4 graphs in the bottom row show residuals.}}\label{TESS_re_ph}
\end{figure*}

\begin{figure}[htbp]
\includegraphics[width=\linewidth]{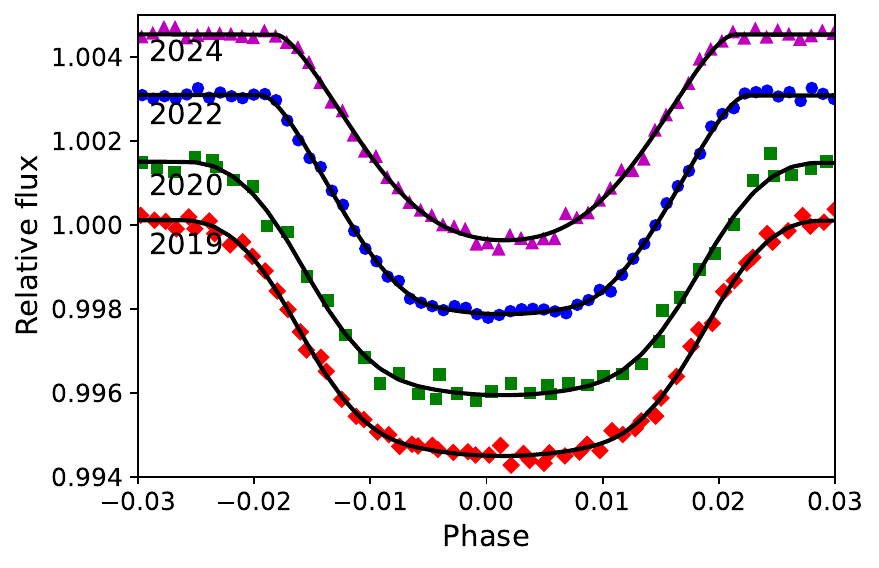}
\caption{Transit part of the phase-folded TESS light curve for the Cloud-free Case. The best-fit light curve models and their corresponding 3-munite binned data (shown as red rhombuses, green squares, blue circles and magenta triangles) are displayed from bottom to top for 2019, 2020, 2022, and 2024, where the latter three curves are offset by +0.0015, +0.003, and +0.0045, respectively.} {Alt text: There is one graph. The x axis shows phase. The y axis shows relative flux.}\label{TESS_re_tr}
\end{figure}

\begin{figure*}[htbp]
\begin{center}
  \includegraphics[width=\linewidth]{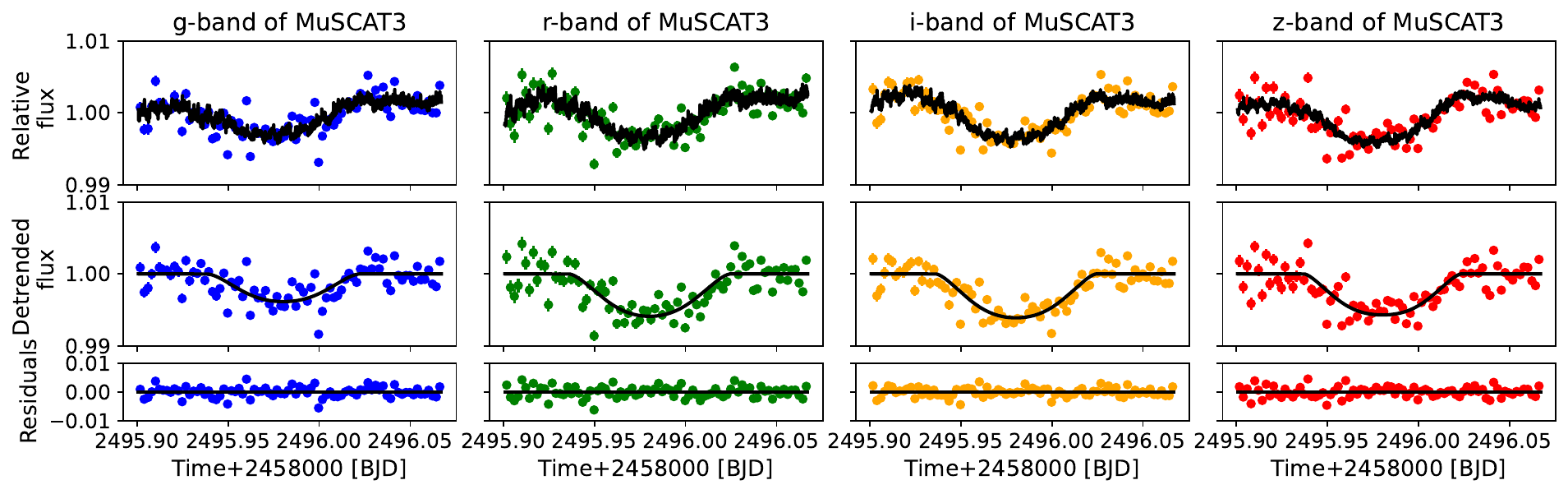}
  \includegraphics[width=\linewidth]{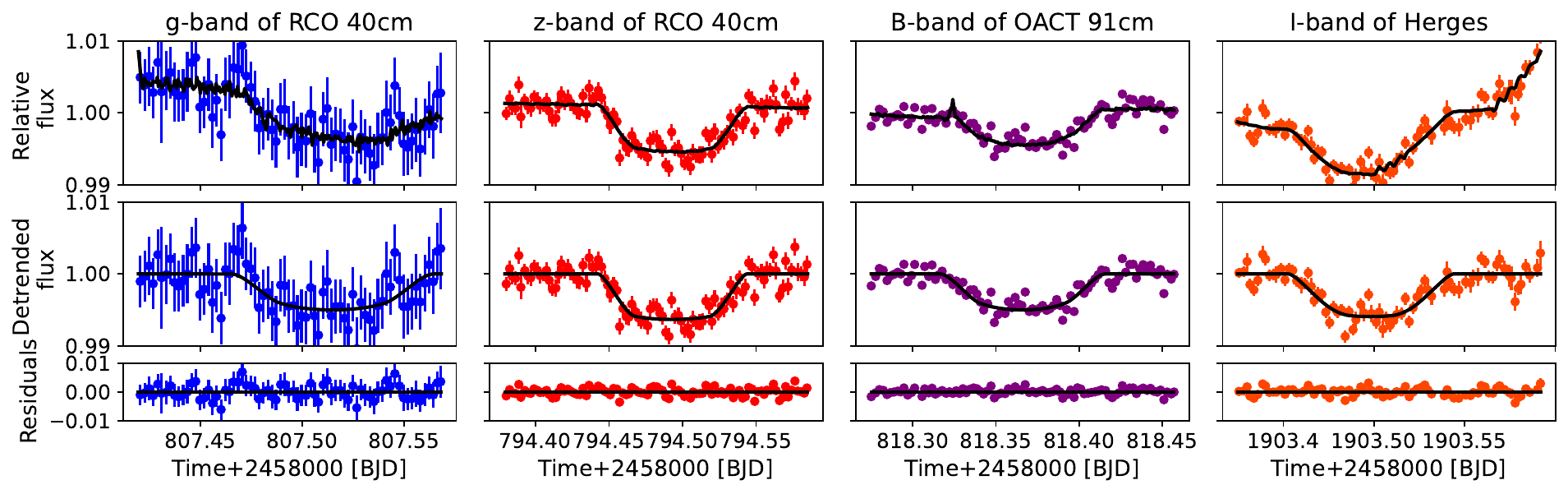}
 \end{center}
\caption{Light curves from MuSCAT3, RCO 40cm, OACT 91cm and SCT in Herges-Hellenburg, showing 3-minute binned data (colored points). The black solid lines represent the models in the Cloud-free Case.
Top panel: Light curves from the photometric observation data. Middle panel: Detrended Light curves. Bottom panel: Residuals between the observed data and model data. {Alt text: 24 line graphs aligned with 6 rows and 4 columns. The x axes of all graphs show time in Barycentric Julian Date. The y axes in top and fourth rows show relative flux. The y axes in second and fifth rows show detrended flux. The y axes in third and bottom rows show relative flux.}}\label{M3_re}
\end{figure*}

The phase-folded TESS light curves with the best-fitting light curve models, transit parts of the phase-folded TESS light curves, and best-fitting light curve models for the ground-based telescopes' data in the Cloud-free Case are presented in figures \ref{TESS_re_ph}, \ref{TESS_re_tr} and \ref{M3_re}, respectively. The calculated values for the Cloud-free Case are presented in table \ref{param_CfC}\footnote{The calculated values for all parameters are shown in e-table 1, which is available only on the online edition as the supplementary data.}.
The residuals in figure \ref{TESS_re_ph} may exhibit small wavy fluctuations of $\sim 100$ ppm. This implies that the possible presence of unaccounted-for components, such as a more intricate variation of $T_D$ and $T_N$ \citep{2021ApJ...915...41M}, or unknown physical components in the light curve data.

We described the results of the Cloud-free Case because the Bayesian Information Criterion (BIC) was minimized in this case. Compared to the minimum BIC, the difference in BIC ($\Delta$BIC) for the Cloudy Case was 43. Moreover, the values of common parameters in the Cloudy Case did not show significant differences from those of the Cloud-free Case. However, reflective clouds are expected to be vaporized due to the high day-side brightness temperature ($T_{D,a}=3630\pm130$ K in 2019, $T_{D,a}=3370^{+180}_{-200}$ K in 2020, $T_{D,a}=3720\pm110$ K in 2022 and $T_{D,a}=3120^{+190}_{-230}$ K in 2024 for the Cloudy Case) \citep{2019A&A...626A.133H}, which is contradictory to the Cloudy Case.
Therefore, we report the values from the Cloudy Case provided in appendix 1. In addition, the posterior distributions from the MCMC analysis for the Cloud-free Case and Cloudy Case are shown in appendix 2.

\begin{table}[htbp]
  \tbl{Parameters of TOI-1355's system from the light curves of TESS and ground-based telescopes for the Cloud-free Case.}{
  \begin{tabular}{lcc}
      \hline
      Parameter for MCMC & Value& Prior \\ 
      \hline
$P_{\mathrm{orb}}$ (days)& $2.1702642^{+7.2\times10^{-6}}_{-7.4\times10^{-6}}$ & $\mathcal{U}(0,100)$\\
$T_{c}$ in 2019\footnotemark[$*$]\footnotemark[$\dag$] & $740.23375\pm0.00020$& $\mathcal{U}(740.0,740.5)$\\
\ (BJD-2458000) &  &  \\
$T_{c}$ in 2020\footnotemark[$*$] & $957.26003\pm0.00028$ & $\mathcal{U}(957.0,957.5)$\\
\ (BJD-2458000) &  &  \\
$T_{c}$ in 2022\footnotemark[$*$]\footnotemark[$\ddag$] & $1853.57908\pm0.00015$ &$\mathcal{U}(1853.5,1854.0)$\\
\ (BJD-2458000) &  & \\
$T_{c}$ in 2024\footnotemark[$*$]\footnotemark[$\S$]& $2396.14475\pm0.00017$ &$\mathcal{U}(2396.0,2396.5)$\\
\ (BJD-2458000) &  & \\
$e$ & $0.2203^{+0.0013}_{-0.0011}$ &$\mathcal{U}(0,1)$\\
$m_{\mathrm{p}}$ ($M_{\mathrm{J}}$) & $5.84^{+0.83}_{-0.78}$ &$\mathcal{U}(0,300)$\\
$M_{\mathrm{s}}$ ($M_{\odot}$) & $2.02\pm0.21$ &  $\mathcal{N}(2.00,0.21)$ \\
$T_{\mathrm{eff}}$ (K)  & $8280^{+360}_{-370}$ & $\mathcal{N}(8660,360)$ \\
$a/R_{\mathrm{s}}$ & $4.589^{+0.049}_{-0.046}$ & $\mathcal{U}(0,100)$\\
$R_{\mathrm{p}}/R_{\mathrm{s}}$ in TESS band& $0.07799\pm0.00041$ & $\mathcal{U}(0.001,0.5)$\\
$R_{\mathrm{p}}/R_{\mathrm{s}}$ in $g$ band & $0.0758^{+0.0085}_{-0.0086}$ & $\mathcal{U}(0.001,0.5)$\\
$R_{\mathrm{p}}/R_{\mathrm{s}}$ in $r$ band & $0.0899^{+0.0093}_{-0.0097}$ & $\mathcal{U}(0.001,0.5)$\\
$R_{\mathrm{p}}/R_{\mathrm{s}}$ in $i$ band & $0.0878\pm0.0080$ & $\mathcal{U}(0.001,0.5)$\\
$R_{\mathrm{p}}/R_{\mathrm{s}}$ in $z$ band & $0.0817^{+0.0034}_{-0.0035}$ & $\mathcal{U}(0.001,0.5)$\\
$R_{\mathrm{p}}/R_{\mathrm{s}}$ in $B$ band  & $0.0750^{+0.0036}_{-0.0038}$ & $\mathcal{U}(0.001,0.5)$\\
$R_{\mathrm{p}}/R_{\mathrm{s}}$ in $I$ band  & $0.0824^{+0.0066}_{-0.0068}$ & $\mathcal{U}(0.001,0.5)$\\
$b$ in 2019\footnotemark[$\dag$] & $0.8531^{+0.0046}_{-0.0047}$ & $\mathcal{U}(0,1.4)$\\
$b$ in 2020 & $0.8623^{+0.0052}_{-0.0058}$ & $\mathcal{U}(0,1.4)$\\
$b$ in 2022\footnotemark[$\ddag$] & $0.9040^{+0.0031}_{-0.0032}$ & $\mathcal{U}(0,1.4)$\\
$b$ in 2024\footnotemark[$\S$] & $0.9275^{+0.0024}_{-0.0025}$ & $\mathcal{U}(0,1.4)$\\
$\omega$ in 2019\footnotemark[$\dag$] (\textdegree) & $182.7\pm3.2$ & $\mathcal{U}(0,360)$\\
$\omega$ in 2020 (\textdegree) & $183.8^{+4.3}_{-4.8}$ & $\mathcal{U}(0,360)$\\
$\omega$ in 2022\footnotemark[$\ddag$] (\textdegree)& $184.5^{+2.5}_{-2.4}$ & $\mathcal{U}(0,360)$\\
$\omega$ in 2024\footnotemark[$\S$] (\textdegree)& $187.3^{+2.4}_{-2.5}$ & $\mathcal{U}(0,360)$\\
$T_{D,a}$ in 2019 (K) &$3730^{+110}_{-120}$&$\mathcal{U}(0,5000)$\\
$T_{D,a}$ in 2020 (K) &$3500\pm120$&$\mathcal{U}(0,5000)$\\
$T_{D,a}$ in 2022 (K) &$3700\pm110$&$\mathcal{U}(0,5000)$\\
$T_{D,a}$ in 2024 (K) &$3440\pm120$&$\mathcal{U}(0,5000)$\\
$T_{N,a}/T_{D,a}$ in 2019 &$0.775^{+0.012}_{-0.014}$&$\mathcal{U}(0,1)$\\
$T_{N,a}/T_{D,a}$ in 2020 &$0.33^{+0.24}_{-0.23}$&$\mathcal{U}(0,1)$\\
$T_{N,a}/T_{D,a}$ in 2022 &$0.748^{+0.012}_{-0.014}$&$\mathcal{U}(0,1)$\\
$T_{N,a}/T_{D,a}$ in 2024 &$0.788^{+0.018}_{-0.029}$&$\mathcal{U}(0,1)$\\
$\delta$ in 2019 (\textdegree) & $-54.5^{+4.1}_{-3.7}$ & $\mathcal{U}(-90,90)$\\
$\delta$ in 2020 (\textdegree) & $-36.3^{+5.8}_{-5.6}$ & $\mathcal{U}(-90,90)$\\
$\delta$ in 2022 (\textdegree) & $-56.0^{+3.3}_{-2.9}$ & $\mathcal{U}(-90,90)$\\
$\delta$ in 2024 (\textdegree) & $-34.4^{+9.1}_{-7.6}$ &  $\mathcal{U}(-90,90)$\\
\hline
Derived Parameter&  Value  & \\
\hline
$a$\footnotemark[$\|$] (AU) & $0.0400\pm0.0012$ & \\
$R_{\mathrm{p}}$\footnotemark[$\|$]\footnotemark[$\#$] ($R_J$) & $1.424^{+0.039}_{-0.041}$ & \\
$P_{\mathrm{orb},0}$ (days) & $2.17026355\pm2.5\times10^{-7}$ & \\
$T_{c, 0}$ (BJD-2458000)  & $740.23374\pm0.00019$ & \\
$db/dt$ (year$^{-1}$) & $ 0.01633^{+0.00081}_{-0.00076}$ & \\
$d\omega/dt$ (\textdegree \ year$^{-1}$)& $0.86^{+0.75}_{-0.69}$ & \\
$T_{N,a}$ in 2019 (K) &$2890^{+100}_{-110}$& \\
$T_{N,a}$ in 2020 (K) &$1170^{+840}_{-800}$&\\
$T_{N,a}$ in 2022 (K) &$2770\pm110$&\\
$T_{N,a}$ in 2024 (K) &$2710^{+130}_{-170}$&\\
\hline
\end{tabular}}\label{param_CfC}
\begin{tabnote}
\footnotemark[$*$] Each $T_{c}$ corresponds to the first transit of TESS observation in each year.\\ 
\footnotemark[$\dag$] These parameters are applied to fit the 2019 TESS data, the RCO 40 cm data, and the OACT 91 cm data.\\
\footnotemark[$\ddag$] These parameters are applied to fit the 2022 TESS data and the data from SCT in Herges-Hallenberg.\\
\footnotemark[$\S$] These parameters are applied to fit the 2024 TESS data and the MuSCAT3 data.\\
\footnotemark[$\|$] These parameters are derived using $R_{\mathrm{s}}$, which is referenced to table 1.\\
\footnotemark[$\#$] This value is derived based on the value of $R_{\mathrm{p}}/R_{\mathrm{s}}$ in TESS band.\\
\end{tabnote}
\end{table}


The $R_{\mathrm{p}}/R_{\mathrm{s}}$ values in seven different filter-bands are consistent within $2 \sigma$.
We also calculated the false positive probability (FPP) that the measured transit depth and orbital period could be explained by astrophysical scenarios other than a planet orbiting the host star using \texttt{TRICERATOPS} (\cite{2020ascl.soft02004G}, \cite{2021AJ....161...24G}). Using the TESS light curve and the contrast curve from WIYN/NESSI speckle imaging, we computed an FPP of $1\times10^{-16}$, which is low enough to formally validate the planet.

We measured planetary mass to be $5.84^{+0.83}_{-0.78} M_J$, planetary radius to be $1.424^{+0.039}_{-0.041} R_J$, orbital period $2.1702642^{+0.0000072}_{-0.0000074}$ days, and eccentricity to be $0.2203^{+0.0013}_{-0.0011}$.
Thus, TOI-1355 b was confirmed to be a massive ($m_{\mathrm{p}}\sim 5.8 M_J$), large ($R_{\mathrm{p}}\sim1.4 R_J$) hot Jupiter with a short orbital period ($P_{\mathrm{orb}}\sim 2.17$ days). This planet orbits an early A-type star ($T_\mathrm{eff} \sim 8280$ K) with a mass of $M_{\mathrm{s}}\sim 2.0 M_{\mathrm{\odot}}$. The orbit exhibits a high eccentricity of $e\sim0.22$, representing one of the highest eccentricities to date among hot Jupiters around hot stars.

\section{Discussion}
\subsection{Circularization of TOI-1355 b}
The high eccentricity of TOI-1355 b implies that this hot Jupiter is experiencing high-eccentricity migration, with the orbit currently undergoing circularization due to the tidal planetary deformation near the periastron. The semi-major axis will reach $0.0381$ AU because the angular momentum of the hot Jupiter's orbit is nearly conserved ($a_{\mathrm{final}}=a(1-e^2)$; \cite{2018ARA&A..56..175D}).
From equation (2) in \citet{2006ApJ...649.1004A}, the circularization timescale is expressed as
\begin{equation}
\label{tcir}
\tau_{\mathrm{cir}}\sim\frac{2Q_{\mathrm{p}} P}{63\pi}\frac{m_{\mathrm{p}}}{M_{\mathrm{s}}}\left(\frac{a}{R_{\mathrm{p}}}\right)^5\frac{(1-e^2)^{13/2}}{1+6e^2}.
\end{equation}
The quality dissipation parameter $Q_{\mathrm{p}}$ is poorly understood. If $Q_{\mathrm{p}}=5\times10^5$ from \citet{2025AJ....170..299K}, the circularization timescale is estimated to be $\tau_{\mathrm{cir}}\sim 3.3\times10^7$ years, which is slightly shorter than the stellar age ($\sim 3.7\times10^8$ years). This suggests that TOI-1355 b may have recently acquired its highly eccentric orbit, or the $Q_{\mathrm{p}}$ value is much higher than $5\times10^5$.

\subsection{Calculation of precise $T_c$ and $P_{\mathrm{orb}}$}
\begin{figure}[htbp]
\includegraphics[width=\linewidth]{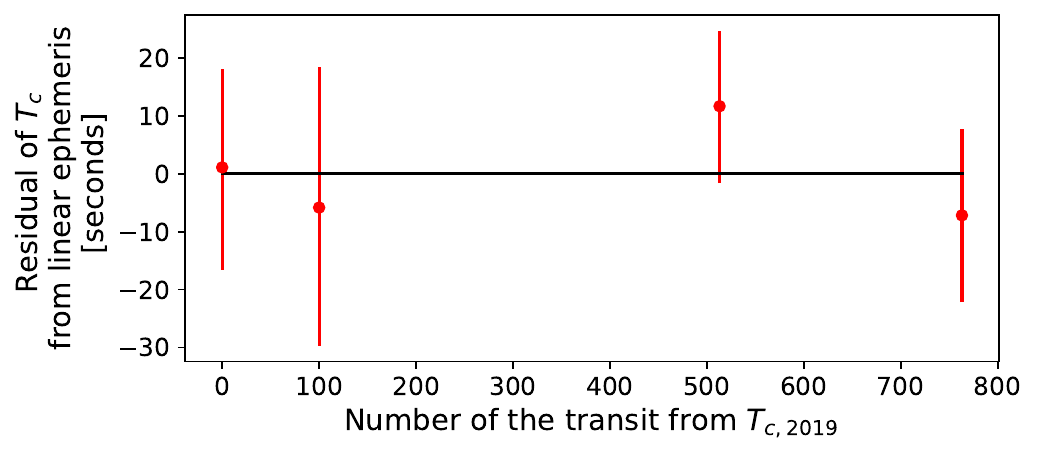}
\caption{The residuals of individual $T_c$ (red points) from the refined linear ephemeris (black line) in the Cloud-free Case. {Alt text: There is one graph. The x axis shows the number of transits from mid-transit in 2019. The y axis shows residuals of individual transit epochs.}}\label{Tchange}
\end{figure}

A precise mid-transit time $T_{c, 0}$ and orbital period $P_{\mathrm{orb},0}$ were derived from a weighted least squares using the following equation:
\begin{equation}
\label{tlin}
T_{c}=T_{c, 0}+nP_{\mathrm{orb},0},
\end{equation}
where $n$ is the epoch number relative to $T_{c,2019}$, and $T_{c}$ denotes the mid-transit time at the $n$th epoch.
Note that $T_{c, 0}$ is not $T_{c}$ of each transit but serves as a reference parameter.
The derived values of $T_{c, 0}$ and $P_{\mathrm{orb,0}}$ are presented in table \ref{param_CfC}, and these values show good agreement with $T_{c,2019}$ and $P_{\mathrm{orb}}$, respectively. Figure \ref{Tchange} shows the residuals of individual Tc from the refined linear ephemeris, indicating no evidence of transit timing variations (TTVs).

\subsection{Nodal Precession}
\begin{figure}[htbp]
   \includegraphics[width=\linewidth]{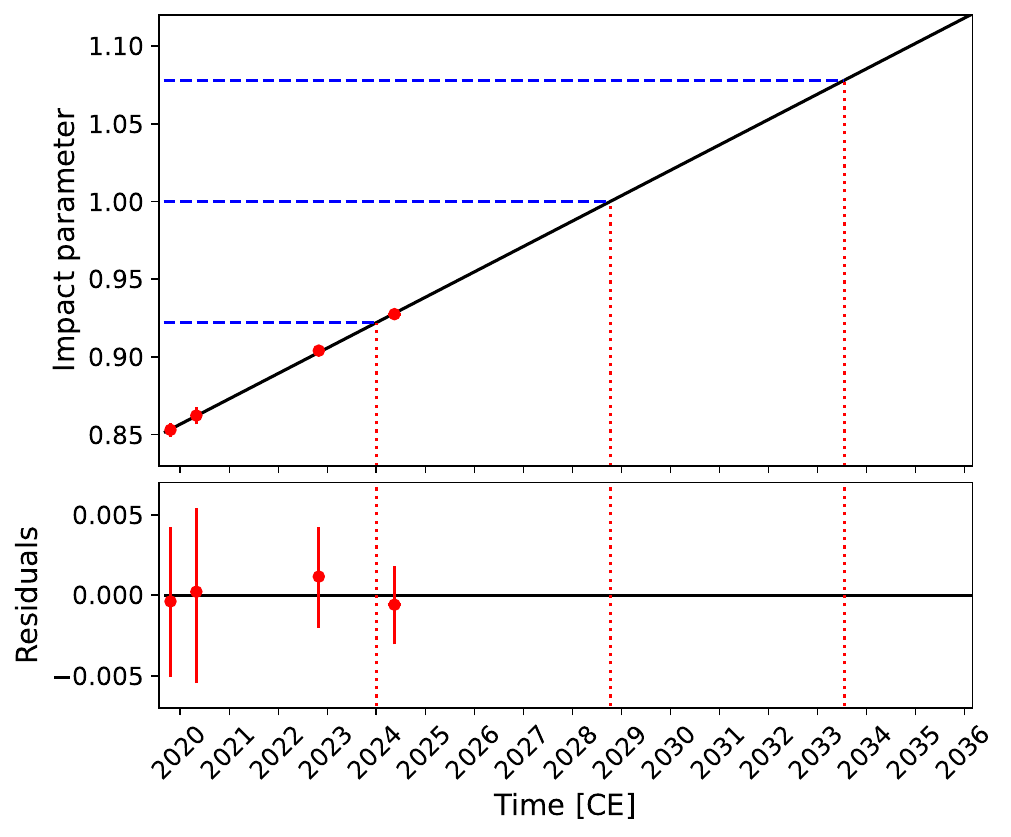}
\caption{Top panel: Change in the impact parameter of TOI-1355 b for the Cloud-free Case. Red points represent the observed data. Black lines show the best-fit models of the change in impact parameter. blue dashed lines mark the positions of $b=1-R_{\mathrm{p}}/R_{\mathrm{s}}$, $b=1$ and $b=1+R_{\mathrm{p}}/R_{\mathrm{s}}$ from bottom to top. The dotted red lines, from left to right, indicate the start of the grazing transit, the epoch at $b=1$, and the end of the grazing transit. Bottom panel: Residuals between the observed data and model {Alt text: 2 graphs. The x axis in all graphs shows the common era. The y axis of top graph shows impact parameter. The y axis of bottom graph shows residuals.}}\label{bchange}
\end{figure}

TOI-1355 b transits the edge of the stellar disk ($b\sim 0.9$). The nodal precession of this hot Jupiter was detected as a change in $b$ over time ($db/dt=0.01633^{+0.00081}_{-0.00076}$ year$^{-1}$ in the Cloud-free Case), although the change in $\omega$ could not be verified. This nodal precession arises under the condition of a rapid rotation of the host star. If $db/dt$ for TOI-1355 b remains constant, the grazing transit began from the beginning of 2024, and the transit geometry completely ceases from the middle of 2033 (see figure \ref{bchange}). 

Here, we derived the relationship between $db/dt$ and the spin-orbit obliquity (projected obliquity $\lambda$ and true obliquity $\psi$) for an eccentric orbit. Differentiating equation (\ref{angle_ip}) with respect to time yields the following equation:
\begin{align}
\label{defip}
\frac{db}{dt}&=-\frac{a}{R_{\rm s}}\frac{1-e^2}{1+e\sin{\omega}}\left(\sin{i_{\rm p}}\frac{di_{\rm p}}{dt}+\frac{e\cos{\omega}\cos{i_{\rm p}}}{1+e\sin{\omega}}\frac{d\omega}{dt} \right) \nonumber \\
&=-b\left(\tan{i_{\rm p}}\frac{di_{\rm p}}{dt}+\frac{e\cos{\omega}}{1+e\sin{\omega}}\frac{d\omega}{dt} \right).
\end{align}
According to \citet{2024PASJ...76..374W}, $i_{\rm p}$ and $\lambda$ can be expressed as
\begin{align}
\label{cosip}
\cos{i_{\rm p}}&=\cos{\psi}\cos{i_{\rm s}}+\sin{\psi}\sin{i_{\rm s}}\cos{\theta} \\
\label{tanlambda}
\tan{\lambda}&=\frac{\sin{\psi}\sin{\theta}}{\sin{\psi}\cos{i_{\rm s}}\cos{\theta}-\cos{\psi}\sin{i_{\rm s}}},
\end{align}
where $\theta$ is the nodal angle (See figure 6 in \citet{2024PASJ...76..374W} for the definitions of the angle). From \citet{2011Ap&SS.331..485I}, we can express $\psi$ as
\begin{align}
\label{psip}
\cos\psi=\cos{i_{\rm p}}\cos{i_{\rm s}}+\sin{i_{\rm p}}\sin{i_{\rm s}}\cos\lambda.
\end{align}
Because the surface gravity is greater than the centrifugal force at the stellar equator \citep{2011Ap&SS.331..485I}, we can impose the range of the $i_s$ to $\timeform{10D}<i_s<\timeform{170D}$ from our values of $R_{\rm s}$, $M_{\rm s}$ and $v\sin i_{\rm s}$.
Using equations (\ref{defip})-(\ref{psip}), we can express $db/dt$ as
\begin{align}
\label{blambda}
\frac{db}{dt}=b\left(\sin{i_{\rm s}}\sin{\lambda}\tan{i_{\rm p}}\frac{d\theta}{dt}-\frac{e\cos{\omega}}{1+e\sin{\omega}}\frac{d\omega}{dt}\right).
\end{align}
In the derivation of equation (\ref{blambda}), $i_s$ is assumed to be constant in time.\footnote{Adopting the moment of inertia coefficient $C \sim 0.3$ from the A-type star WASP-33 (\cite{2011Ap&SS.331..485I}), the ratio of the planetary orbital angular momentum ($L_{\rm p}=2\pi m_{\rm p}a^2\sqrt{1-e^2}/P_{\rm orb}$) to the stellar spin angular momentum ($L_{\rm{s}}=2\pi CM_{\rm s}R_{\rm s}^2/P_{\rm spin}\ge CM_{\rm s}R_{\rm s}v\sin i_{\rm s}$) is estimated as $L_{\rm{p}}/L_{\rm{s}}\lesssim0.1$. Thus, the stellar spin axis remains nearly stationary, allowing us to treat $i_s$ as constant.}
Following \citet{2011PhRvD..84l4001I}, $d\theta/dt$ can be given by
\begin{align}
\label{dthdt}
\frac{d\theta}{dt}=-\frac{3\pi J_2 R_{\rm s}^2}{P_{\rm orb}a^2(1-e^2)^2}\cos\psi,
\end{align}
where $J_2$ is the stellar quadrupole moment. Here, based on \citet{2009ApJ...698.1778R}, $J_2$ can be expressed as follows:
\begin{align}
\label{J2}
J_2=\frac{k_{2,\rm s}R_{\rm s}^3}{3a^3}\left(\frac{P_{\rm orb}}{P_{\rm spin}}+\frac{3m_{\rm p}}{2M_{\rm s}}\right),
\end{align}
where $k_{2,\rm s}$ is the Love number of the host star and $P_{\rm spin} (=2\pi R_{\rm s} \sin i_{\rm s}/ (v\sin i_{\rm s}))$ is the stellar rotation speed. Regarding $d\omega/dt$, as shown in \citet{fabrycky2010non}, it can be described as follows:
\begin{align}
\label{dwdt}
\frac{d\omega}{dt}&=\left(\frac{d\omega}{dt}\right)_{\rm G}+\left(\frac{d\omega}{dt}\right)_{\rm T}+\left(\frac{d\omega}{dt}\right)_{\rm R},
\end{align}
where 
\begin{align}
\label{dwdt_G}
\left(\frac{d\omega}{dt}\right)_{\rm G}&=\frac{24\pi^3a^2}{c^2P_{\rm orb}^3(1-e^2)}  \\ 
\label{dwdt_T}
\left(\frac{d\omega}{dt}\right)_{\rm T}&=\frac{15\pi k_{2,\rm p}M_{\rm s}R_{\rm p}^5}{P_{\rm orb}m_{\rm p}a^5}\frac{1+(3/2)e^2-(1/8)e^4}{(1-e^2)^5} \\
\label{dwdt_R}
\left(\frac{d\omega}{dt}\right)_{\rm R}&=\frac{\pi k_{2,\rm s}P_{\rm orb}R_{\rm s}^5(1+m_{\rm p}/M_{\rm s})}{P_{\rm spin}^2 a^5 (1-e^2)^2}\frac{5\cos^2\psi-1}{4}.
\end{align}
Here, $c$ is the speed of light, equation (\ref{dwdt_G}) shows the periastron advance rate due to the relativistic effect, equation (\ref{dwdt_T}) described the apsidal motion caused by the planetary tidal distortion, and equation (\ref{dwdt_R}) is the apsidal motion due to the stellar rotational distortion. Under the assumption that $m_{\rm p}\ll M_{\rm s}$, equations (\ref{blambda})-(\ref{dwdt_R}) yield the following expression for $db/dt$:
\begin{align}
\label{blambda_2}
\frac{db}{dt}&=-\frac{4b\sin{i_{\rm s}}\sin{\lambda}\tan{i_{\rm p}}\cos\psi}{5\cos^2 \psi-1}\left(\frac{d\omega}{dt}\right)_{\rm R}-\frac{be\cos{\omega}}{1+e\sin{\omega}}\frac{d\omega}{dt} \nonumber \\
&=-\frac{4b\sin{i_{\rm s}}\sin{\lambda}\tan{i_{\rm p}}\cos\psi}{5\cos^2 \psi-1}\left\{\frac{d\omega}{dt}-\left(\frac{d\omega}{dt}\right)_{\rm T}-\left(\frac{d\omega}{dt}\right)_{\rm R}\right\} \nonumber \\
&-\frac{be\cos{\omega}}{1+e\sin{\omega}}\frac{d\omega}{dt}.
\end{align}

By calculating $\lambda$ and $\psi$ from equations (\ref{psip}), (\ref{dwdt_G}), (\ref{dwdt_T}) and (\ref{blambda_2}) using $i_{\rm s}$ within the range of $\timeform{10D}<i_{\rm s}<\timeform{170D}$ and our derived values of $db/dt$, $d\omega/dt$, $b$, $\omega$, $a/R_{\rm s}$, $R_{\rm p}/R_{\rm s}$ (in TESS band), $a$, $e$, $P_{\rm orb}$, $M_{\rm s}$ and $m_{\rm p}$ within a $1\sigma$-confidence level, we constrained their ranges to $\timeform{-76D}<\lambda<\timeform{-5D}$ or $\timeform{104D}<\lambda<\timeform{175D}$, and $\timeform{7D}<\psi<\timeform{76D}$ or $\timeform{104D}<\psi<\timeform{173D}$. Additionally, the ranges of $(d\omega/dt)_{\rm G}$ and $(d\omega/dt)_{\rm T}$ were calculated to $\timeform{0.D080}\ \rm{year}^{-1}<(d\omega/dt)_{\rm G}<\timeform{0.D091}\ \rm{year}^{-1}$ and $\timeform{0.D15}\ \rm{year}^{-1}<(d\omega/dt)_{\rm T}<\timeform{0.D28}\ \rm{year}^{-1}$ from equations (\ref{dwdt_G}) and (\ref{dwdt_T}). In this calculation, we assumed $k_{2, \rm p}=0.63$ from the value of a hot Jupiter WASP-12 b in \citet{2026AJ....171..246B}. Currently, $\psi$ is only loosely constrained. Previous studies have detected nodal precession in four hot Jupiters: Kepler-13Ab \citep{2012MNRAS.421L.122S,2014MNRAS.437.1045S,2018AJ....155...13H}, WASP-33b \citep{2015ApJ...810L..23J,2020PASJ...72...19W,2022MNRAS.512.4404W,2022ApJ...931..111S}, KELT-9b \citep{2022ApJ...931..111S} and TOI-1518b \citep{2024PASJ...76..374W}. These hot Jupiters revolve around rapidly rotating hot stars in near-polar orbits ($\timeform{50D}<\psi<\timeform{130D}$).
To further tighten the constraints on $\psi$ and verify whether the orbit of TOI-1355 b is near-polar like these planetary systems, the value of $\lambda$ should be measured through Doppler tomography \citep{2010MNRAS.407..507C}. Moreover, using equations (\ref{dwdt})-(\ref{dwdt_R}), we calculated the Love number of TOI-1355 $\log k_{2,\rm s}=-1.83\pm0.40$, which agrees with that of the Sun ($\log k_{2,\rm s}=-1.52$; \cite{1995AAS..114..549C}) within $1\sigma$.



\subsection{Atmosphere of TOI-1355 b}
In the Cloud-free Case, on the other hand, the day-side brightness temperature is above 3000 K.
The equilibrium temperature of TOI-1355 at a distance of $a$ from the host star can be given by
\begin{equation}
\label{teq}
T_{\mathrm{eq}, a} = \left(\frac{1}{4}\right)^{1/4}T_{\mathrm{eff}}\sqrt{\frac{R_{\mathrm{s}}}{a}},
\end{equation}
yielding $T_{\mathrm{eq}, a}=2740\pm120$ K, which is lower than the brightness temperature. This could be caused by factors such as thermal inversions due to the presence of TiO and VO (\cite{2003ApJ...594.1011H}) or inefficient heat redistribution efficiency \citep{2011ApJ...729...54C}. 
To investigate the existence of molecules that cause thermal inversion on TOI-1355 b, further atmospheric follow-up observations are needed for this hot Jupiter.
The derived equilibrium temperature is higher than the night-side brightness temperature in 2020 ($T_{N,a}=1170^{+840}_{-800}$ K in 2020), while it is consistent with those in 2019, 2022 and 2024 within $1\sigma$ ($T_{N,a}=2890^{+100}_{-110}$ K in 2019, $T_{N,a}=2770\pm110$ K in 2022, and $T_{N,a}=2710\pm110$ K in 2024). Regarding $T_{N,a}$ in 2022 and 2024, they are marginally higher than $T_{\mathrm{eq}, a}$. Tidal heating could be a contributing factor to the higher night-side temperature \citep{2010A&A...516A..64L}, but the underlying cause requires further investigation in future studies.
Moreover, the phase shifts in 2020 and 2022 ($\delta=\timeform{-36.D3}^{+5.8}_{-5.6}$ in 2020 and  $\delta=\timeform{-56.D0}^{+3.3}_{-2.9}$ in 2022) are inconsistent beyond $3 \sigma$. One possibility for this difference could be a variable atmosphere on TOI-1355 b \citep{2024ApJS..270...34C}.

\subsection{TOI-1355 Variability Analysis}
A-type main-sequence stars such as TOI-1355 are in the classical instability strip of the H-R diagram and can thus become unstable to $\delta$ Scuti pulsations, which are driven by the $\kappa$ mechanism \citep{1979PASP...91....5B}. $\delta$ Scuti stars pulsate with typical frequencies of $\sim 10-100$ cycles day$^{-1}$ and amplitudes of $\sim$1\,mmag, and thus can significantly affect the modeling of transits \citep{2021AJ....162..204H}. Pulsations can also be used to refine the fundamental parameters of the host star, such as stellar age (e.g. \cite{2021MNRAS.502.1633M}, \cite{2024ApJ...960...94P}) and line of sight inclinations, which can inform planetary architectures (e.g. \cite{2019A&A...627A..28Z}, \cite{2024AJ....168...13S}).

\begin{figure}[htbp]
\begin{center}
\includegraphics[width=\linewidth]{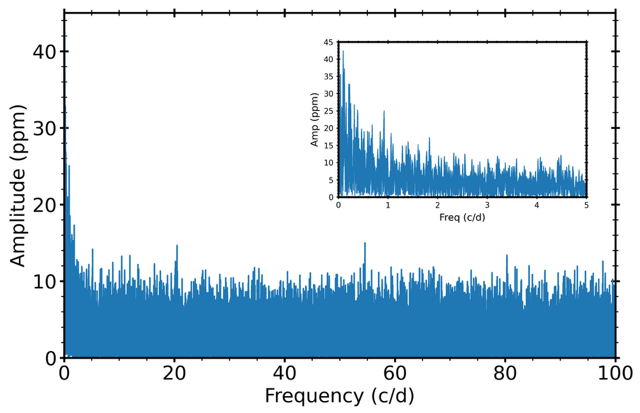}
\end{center}
\caption{Amplitude spectrum of the combined TESS light curve after removing the transit and phase curve model. {Alt text: There is one graph including an inset with the low frequency region. The x axis shows frequency. The y axis shows flux.}}
\label{fig:ampspec}
\end{figure}

To search for pulsations, we calculated an amplitude spectrum of the combined TESS light curves after removing the transit and phase curve model described in section 4. Figure \ref{fig:ampspec} shows no pulsations down to a $3 \sigma$ limit of $\sim 30$\,ppm, well below the typical pulsation amplitude of $\delta$ Scuti stars \citep{2025MNRAS.542.2866M}. The signal at very low frequencies ($<5$ cycles day$^{-1}$) can be attributed to small variations in the residual light curve after removing the planetary signal due to imperfect modeling of the transits. For example, the peak at $\sim0.92$ cycles day$^{-1}$ corresponds to twice to the orbital period of the planet. \citet{2024ApJ...972..137G} demonstrated that the pulsator fraction for the approximate color-magnitude diagram of TOI-1355 is $\sim 50$\%, increasing steeply for rapid rotators with $v\sin i_{\rm s}\gtrsim 100$ km s$^{-1}$. We conclude that the non-detection of pulsations in TOI-1355 is consistent with statistics of known $\delta$ Scuti stars, and that stellar variability has a negligible effect on the transit analysis described in section 4.

\section{Conclusion}
We discovered a hot super-Jupiter ($m_{\mathrm{p}}\sim5.8M_J$), TOI-1355 b, with a radius $R_{\mathrm{p}}\sim1.5R_J$, orbiting an early A-type star. This is the one of the few eccentric hot Jupiters around hot stars ($e\sim0.22$), making TOI-1355 b an intriguing target for investigating high-eccentricity migration.
We also detected the planet transiting the edge of the stellar disk ($b\sim 0.9$) and the nodal precession of TOI-1355 b, where the impact parameter increases by $db/dt\sim0.016$ year$^{-1}$. We estimate that the transit will cease completely by the middle of 2033.

Our analysis of the phase curve suggests that molecules causing thermal inversions would likely exist on TOI-1355 b.
Atmospheric observations using the high-resolution spectrographs (e.g. Subaru/HDS; \cite{2002PASJ...54..855N}) and space telescopes such as Hubble Space Telescope (HST) and James Webb Space Telescope (JWST) will clarify whether TOI-1355 b has particles responsible for thermal inversions.

We also found time variations in the phase shift. 
To comprehend how these parameters change, we should continue to obtain the out-of-transit light curve with TESS and JWST for a long period, which would help elucidate the atmospheric behavior of this hot Jupiter.
Finally, we note that these observations should be conducted swiftly by the middle of 2033, after which the disk of TOI-1355 b will not transit.

\section*{Supplementary data} 
The following supplementary data is available in the online version of this article.
E-tables 1 and 2, and e-figures 1 and 2.

\begin{ack}
This work is partly supported by JSPS KAKENHI Grant Numbers JP24H00017, JP24K00689, JP25K17450, JP25KJ0091, JP25KJ1036, JST SPRING Grant Number JPMJSP2108, and JSPS Bilateral Program Number JPJSBP120249910.
M.T. is supported by JSPS KAKENHI Grant Number JP24H00242.
D.H. acknowledges support from the National Aeronautics and Space Administration (80NSSC22K0781).
This study makes use of data obtained through the observation programs at the 3.8-m Seimei telescope, 23B-N-CN01 in the observing time provided by NAOJ and 23B-K-0008 in the observing time of Kyoto University.
The GAOES-RV project started as a collaboration between Gunma
Astronomical Observatory and Institute of Science Tokyo. Based on the contract signed between the two parties, GAOES-RV is lent to Institute of Science Tokyo and operated at the Seimei Telescope.
This paper is based on observations made with the MuSCAT3 instruments, developed by the Astrobiology Center (ABC) in Japan, the University of Tokyo, and Las Cumbres Observatory (LCOGT). MuSCAT3 was developed with financial support by JSPS KAKENHI (JP18H05439) and JST PRESTO (JPMJPR1775), and is located at the Faulkes Telescope North on Maui, HI (USA), operated by LCOGT.
Funding for the TESS mission is provided by NASA's Science Mission Directorate.
We acknowledge the use of TESS High Level Science Products (HLSP) produced by TESS Light Curves From Full Frame Images (TESS-SPOC).
Resources supporting this work were provided by the NASA High-End Computing (HEC) Program through the NASA Advanced Supercomputing (NAS) Division at Ames Research Center for the production of the SPOC data products.
We acknowledge the use of public TESS data from pipelines at the TESS Science Office and at the TESS Science Processing Operations Center.
This paper includes data collected by the TESS mission that are publicly available from the Mikulski Archive for Space Telescopes (MAST).
This research has made use of the Exoplanet Follow-up Observation Program website, which is operated by the California Institute of Technology, under contract with the National Aeronautics and Space Administration under the Exoplanet Exploration Program.
This work has made use of the Vienna Atomic Line Database (VALD) database, operated at Uppsala University, the Institute of Astronomy RAS in Moscow, and the University of Vienna.
\end{ack}


\section*{Data availability} 
The reduced data from Seimei/GAOES-RV, LCO2m/MuSCAT3, RCO/40cm telescope, OACT/91cm telescope and SCT in Herges-Hallenberg will be shared on reasonable request to the corresponding author. The TESS datasets are available from the Mikulski Archive for Space Telescopes (MAST; https://archive.stsci.edu/missions-and-data/tess).

\appendix
\section{Results from TESS and Ground-based Observations in Cloudy Case}
The calculated values for the data from the TESS and ground-based observations in the Cloudy Case are presented in table \ref{param_CC}.\footnote{The calculated values for all parameters are shown in e-table 2, which is available only on the online edition as the supplementary data.}
\begin{table}[htbp]
  \tbl{Parameters of TOI-1355's system from the light curves of TESS and ground-based telescopes for the Cloudy Case.}{
  \begin{tabular}{lcc}
      \hline
      Parameter for MCMC & Value& Prior \\ 
      \hline
$P_{\mathrm{orb}}$ (days)& $2.1702638^{+7.5\times10^{-6}}_{-7.2\times10^{-6}}$ & $\mathcal{U}(0,100)$\\
$T_{c}$ in 2019\footnotemark[$*$]\footnotemark[$\dag$] & $740.23377^{+0.00020}_{-0.00021}$& $\mathcal{U}(740.0,740.5)$\\
\ (BJD-2458000) &  &  \\
$T_{c}$ in 2020\footnotemark[$*$] & $957.26007\pm0.00028$ & $\mathcal{U}(957.0,957.5)$\\
\ (BJD-2458000) &  &  \\
$T_{c}$ in 2022\footnotemark[$*$]\footnotemark[$\ddag$] & $1853.57908\pm0.00016$ &$\mathcal{U}(1853.5,1854.0)$\\
\ (BJD-2458000) &  & \\
$T_{c}$ in 2024\footnotemark[$*$]\footnotemark[$\S$]& $2396.14476\pm0.00018$ &$\mathcal{U}(2396.0,2396.5)$\\
\ (BJD-2458000) &  & \\
$e$ & $0.2202^{+0.0013}_{-0.0011}$ &$\mathcal{U}(0,1)$\\
$m_{\mathrm{p}}$ ($M_{\mathrm{J}}$) & $6.69^{+0.90}_{-0.87}$ &$\mathcal{U}(0,300)$\\
$M_{\mathrm{s}}$ ($M_{\odot}$) & $2.02\pm0.21$ &  $\mathcal{N}(2.00,0.21)$ \\
$T_{\mathrm{eff}}$ (K) & $8490^{+370}_{-380}$ & $\mathcal{N}(8660,360)$ \\
$a/R_{\mathrm{s}}$ & $4.588^{+0.049}_{-0.046}$ & $\mathcal{U}(0,100)$\\
$R_{\mathrm{p}}/R_{\mathrm{s}}$ in TESS Band& $0.07794^{+0.00040}_{-0.00039}$ & $\mathcal{U}(0.001,0.5)$\\
$R_{\mathrm{p}}/R_{\mathrm{s}}$ in $g$ Band & $0.0759^{+0.0085}_{-0.0089}$ & $\mathcal{U}(0.001,0.5)$\\
$R_{\mathrm{p}}/R_{\mathrm{s}}$ in $r$ Band & $0.0894^{+0.0097}_{-0.0096}$ & $\mathcal{U}(0.001,0.5)$\\
$R_{\mathrm{p}}/R_{\mathrm{s}}$ in $i$ Band & $0.0879^{+0.0081}_{-0.0082}$ & $\mathcal{U}(0.001,0.5)$\\
$R_{\mathrm{p}}/R_{\mathrm{s}}$ in $z$ Band & $0.0814^{+0.0034}_{-0.0035}$ & $\mathcal{U}(0.001,0.5)$\\
$R_{\mathrm{p}}/R_{\mathrm{s}}$ in $B$ Band  & $0.0750^{+0.0036}_{-0.0038}$ & $\mathcal{U}(0.001,0.5)$\\
$R_{\mathrm{p}}/R_{\mathrm{s}}$ in $I$ Band  & $0.0824^{+0.0066}_{-0.0069}$ & $\mathcal{U}(0.001,0.5)$\\
$b$ in 2019\footnotemark[$\dag$] & $0.8520^{+0.0049}_{-0.0050}$ & $\mathcal{U}(0,1.4)$\\
$b$ in 2020 & $0.8611^{+0.0055}_{-0.0060}$ & $\mathcal{U}(0,1.4)$\\
$b$ in 2022\footnotemark[$\ddag$] & $0.9039^{+0.0031}_{-0.0034}$ & $\mathcal{U}(0,1.4)$\\
$b$ in 2024\footnotemark[$\S$] & $0.9272^{+0.0025}_{-0.0027}$ & $\mathcal{U}(0,1.4)$\\
$\omega$ in 2019\footnotemark[$\dag$] (\textdegree) & $182.1^{+3.4}_{-3.3}$ & $\mathcal{U}(0,360)$\\
$\omega$ in 2020 (\textdegree) & $183.1^{+4.7}_{-5.3}$ & $\mathcal{U}(0,360)$\\
$\omega$ in 2022\footnotemark[$\ddag$] (\textdegree)& $184.4^{+2.6}_{-2.4}$ & $\mathcal{U}(0,360)$\\
$\omega$ in 2024\footnotemark[$\S$] (\textdegree)& $187.2^{+2.5}_{-2.6}$ & $\mathcal{U}(0,360)$\\
$A_{g}$ in 2019 &$0.30^{+0.14}_{-0.16}$& $\mathcal{U}(0,2)$\\
$A_{g}$ in 2020 & $0.28^{+0.20}_{-0.18}$ & $\mathcal{U}(0,2)$\\
$A_{g}$ in 2022 & $0.037^{+0.054}_{-0.028}$ & $\mathcal{U}(0,2)$\\
$A_{g}$ in 2024 & $0.422^{+0.068}_{-0.101}$ & $\mathcal{U}(0,2)$\\
$T_{D,a}$ in 2019 (K) &$3630\pm130$&$\mathcal{U}(0,5000)$\\
$T_{D,a}$ in 2020 (K) &$3370^{+180}_{-200}$&$\mathcal{U}(0,5000)$\\
$T_{D,a}$ in 2022 (K) &$3720\pm110$&$\mathcal{U}(0,5000)$\\
$T_{D,a}$ in 2024 (K) &$3120^{+190}_{-230}$&$\mathcal{U}(0,5000)$\\
$T_{N,a}/T_{D,a}$ in 2019 &$0.773^{+0.014}_{-0.016}$&$\mathcal{U}(0,1)$\\
$T_{N,a}/T_{D,a}$ in 2020 &$0.34\pm0.24$&$\mathcal{U}(0,1)$\\
$T_{N,a}/T_{D,a}$ in 2022 &$0.745^{+0.012}_{-0.014}$&$\mathcal{U}(0,1)$\\
$T_{N,a}/T_{D,a}$ in 2024 &$0.838^{+0.029}_{-0.024}$&$\mathcal{U}(0,1)$\\
$\delta$ in 2019 (\textdegree) & $-72\pm11$ & $\mathcal{U}(-90,90)$\\
$\delta$ in 2020 (\textdegree) & $-50^{+12}_{-17}$ & $\mathcal{U}(-90,90)$\\
$\delta$ in 2022 (\textdegree)& $-58.4^{+3.8}_{-3.9}$ & $\mathcal{U}(-90,90)$\\
$\delta$ in 2024 (\textdegree) & $-72^{+19}_{-13}$ &  $\mathcal{U}(-90,90)$\\
\hline
Derived Parameter& Value & \\
\hline
$a$\footnotemark[$\|$] (AU) & $0.0400\pm0.0012$ & \\
$R_{\mathrm{p}}$\footnotemark[$\|$]\footnotemark[$\#$] ($R_J$) & $1.423^{+0.039}_{-0.041}$ & \\
$P_{\mathrm{orb},0}$ (days) & $2.17026353\pm2.6\times10^{-7}$ & \\
$T_{c, 0}$ (BJD-2458000)  & $740.23377\pm0.00019$ &\\
$db/dt$ (year$^{-1}$) & $ 0.01655^{+0.00085}_{-0.00079}$ & \\
$d\omega/dt$ (\textdegree \ year$^{-1}$)& $1.00^{+0.78}_{-0.75}$ & \\
$T_{N,a}$ in 2019 (K) &$2810^{+120}_{-130}$&\\
$T_{N,a}$ in 2020 (K) &$1140^{+840}_{-780}$&\\
$T_{N,a}$ in 2022 (K) &$2770\pm110$&\\
$T_{N,a}$ in 2024 (K) &$2620^{+130}_{-160}$&\\
\hline
\end{tabular}}\label{param_CC}
\begin{tabnote}
\footnotemark[$*$] Each $T_{c}$ corresponds to the first transit of TESS observation in each year.\\ 
\footnotemark[$\dag$] These parameters are applied to fit the 2019 TESS data, the RCO 40 cm data, and the OACT 91 cm data.\\
\footnotemark[$\ddag$] These parameters are applied to fit the 2022 TESS data and the data from SCT in Herges-Hallenberg.\\
\footnotemark[$\S$] These parameters are applied to fit the 2024 TESS data and the MuSCAT3 data.\\
\footnotemark[$\|$] These parameters are derived using $R_{\mathrm{s}}$, which is referenced to table 1.\\
\footnotemark[$\#$] This value is derived based on the value of $R_{\mathrm{p}}/R_{\mathrm{s}}$ in TESS band.\\
\end{tabnote}
\end{table}

\section{MCMC Results of Photometric Measurements by TESS and Ground-based Telescope}
\begin{figure*}[htbp]
\begin{center}
  \includegraphics[width=105mm]{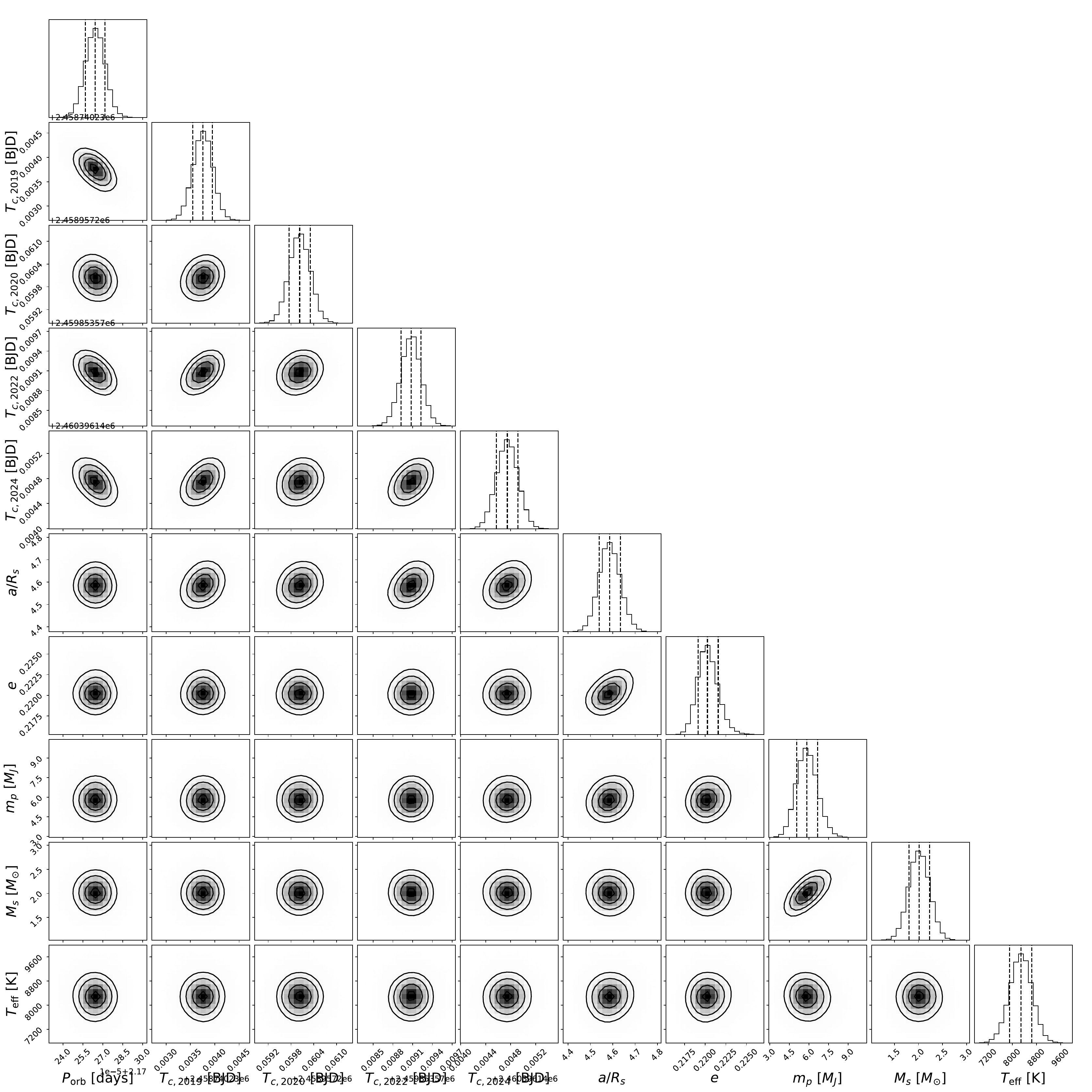}
 \end{center}
\caption{Corner plots of the free parameters in table \ref{param_CfC} for the Cloud-free Case. These plots were generated using corner.py \citep{corner}. {Alt text: Posterior plots of the calculated parameters in MCMC.}}\label{corner_CfC}
\end{figure*}

\begin{figure*}[htbp]
\begin{center}
  \includegraphics[width=105mm]{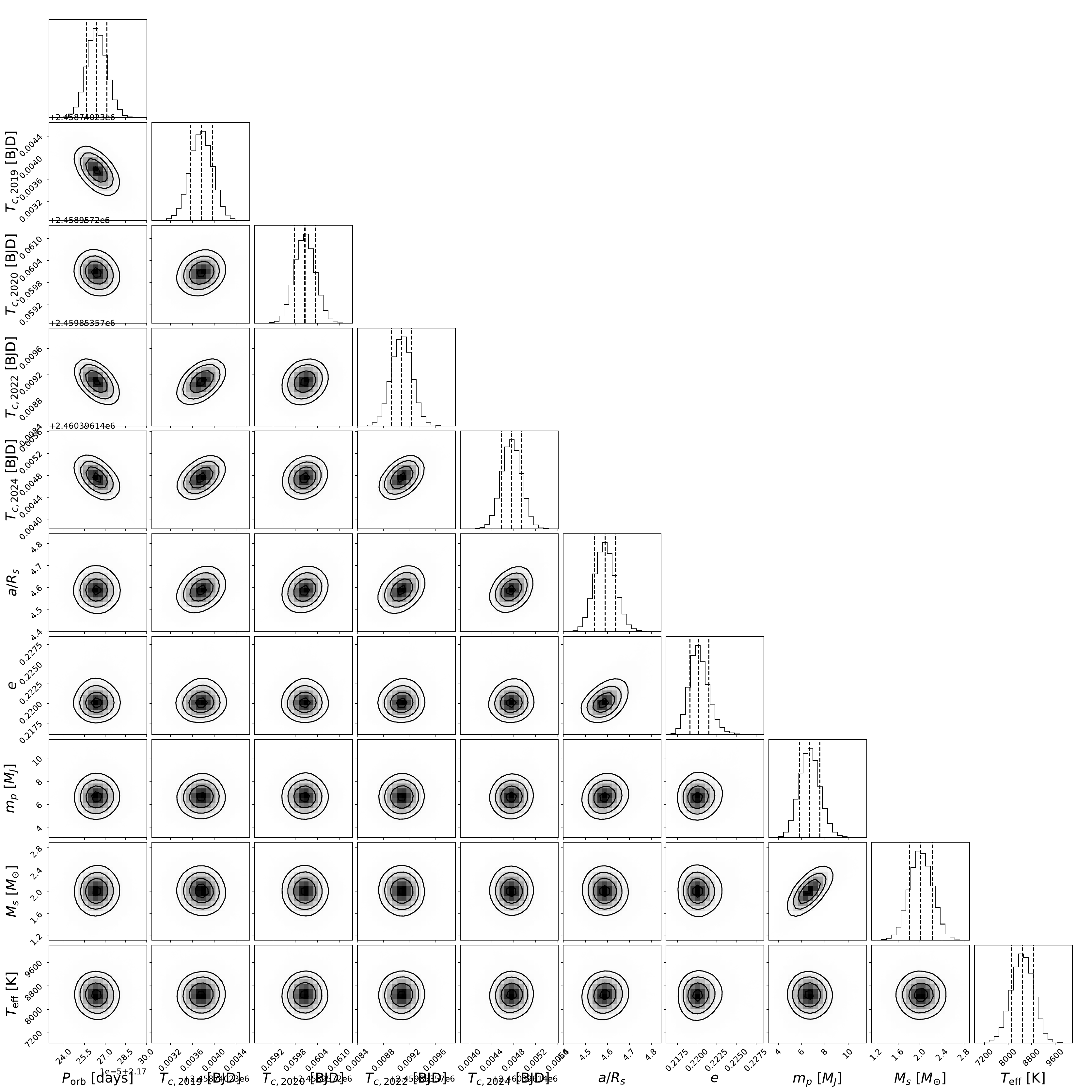}
 \end{center}
\caption{Corner plots of the free parameters in table \ref{param_CC} for the Cloudy Case. These plots were generated in the same way as figure \ref{corner_CfC}. {Alt text: Posterior plots of the calculated parameters in MCMC.}}\label{corner_CC}
\end{figure*}

The corner plots of the posteriors calculated by MCMC are shown for the Cloud-free Case and Cloudy Case in figure \ref{corner_CfC} and figure \ref{corner_CC}, respectively.\footnote{We display a part of the corner plots in figures \ref{corner_CfC} and \ref{corner_CC}. The corner plots for all parameters are shown in e-figures 1 and 2 for the Cloud-free Case and Cloudy Case, respectively, which are available only on the online edition as supplementary data.}

\bibliographystyle{apj}
\bibliography{reference}


\end{document}